\documentclass[structabstract]{aa}			

\usepackage{graphicx}
\usepackage{color}
\usepackage{txfonts}
\usepackage{longtable,lscape}
\usepackage{natbib}

\begin{document}
%

   \title{A 3D view of the outflow in the Orion Molecular Cloud 1 (\object{OMC-1})\thanks{Based on observations obtained at the Canada-France-Hawaii Telescope (CFHT) which is operated by the National Research Council of Canada, the Institut National des Sciences de lÕUnivers of the Centre National de la Recherche Scientifique of France, and the University of Hawaii.}}


   \author{H. D. Nissen
          \inst{1}
          \and
          N.J. Cunningham\inst{2}
          \and 
          M. Gustafsson\inst{3}
          \and 
          J. Bally\inst{4}
          \and
          J.-L. Lemaire\inst{5,}\thanks{Visiting astronomer at the Canada-France-Hawaii Telescope, Mauna Kea, Hawaii.}
          \and
          C. Favre\inst{1}    
          \and
          D. Field\inst{1,}{$^{\star\star}$}
          }
   \institute{Department of Physics and Astronomy, University of \AA rhus, Ny Munkegade, DK-8000 \AA rhus C, Denmark\\
              \email{henrik.dahl.nissen@gmail.com, dfield@phys.au.dk, favre@phys.au.dk}
         \and
            Department of Physics and Astronomy, University of Nebraska-Lincoln 116 Brace Laboratory, Lincoln, NE 68588-0111\\
             \email{ncunningham2@unl.edu}       
         \and
	  Max Planck Institute for Astronomy, K\"{o}nigstuhl 17, 69117 Heidelberg, Germany\\
             \email{gustafsson@mpia.de}       
         \and
            Center for Astrophysics and Space Astronomy, University of Colorado, Boulder, CO 80309\\
             \email{John.Bally@colorado.edu}       
         \and         
             LAMAp/LERMA, UMR8112 du CNRS, de l'Observatoire de Paris et de l'Universit\'e de Cergy Pontoise, 95031 Cergy Pontoise Cedex, France\\
             \email{jean-louis.lemaire@obspm.fr}              
             }

   \date{Received Month Day, 2011; accepted Month Day, 2012}




  \abstract
   {Stars whose mass is an order of magnitude greater than the Sun play a prominent role in the evolution of galaxies, exploding as supernovae, triggering bursts of star formation and spreading heavy elements about their host galaxies. A fundamental aspect of star formation is the creation of an outflow. The fast outflow emerging from a region associated with massive star formation in the Orion Molecular Cloud 1 (\object{OMC-1}), located behind the Orion Nebula, appears to have been set in motion by an explosive event.}
   {Here we study the structure and dynamics of outflows in \object{OMC-1}. We combine radial velocity and proper motion data for near-IR emission of molecular hydrogen to obtain the first 3-dimensional (3D) structure of the \object{OMC-1} outflow. Our work illustrates a new diagnostic tool for studies of star formation that will be exploited in the near future with the advent of high spatial resolution spectro-imaging in particular with data from the Atacama Large Millimeter Array (ALMA).}
   {We use published radial and proper motion velocities obtained from the shock-excited vibrational emission  in the H$\rm_{2}$ v~=~1--0 S(1) line at 2.122~$\mu$m obtained with the GriF instrument on the Canada-France-Hawaii Telescope, the Apache Point Observatory, the Anglo-Australian Observatory and the Subaru Telescope. }
   {These data give the 3D velocity of ejecta yielding a 3D reconstruction of the outflows. This allows one to view the material from different vantage points in space giving considerable insight into the geometry. Our analysis indicates that the ejection occurred $\la$~720~years ago from a distorted ring-like structure of $\sim$15$\arcsec$ (6000~AU) in diameter centered on the proposed point of close encounter of the stars BN, source I and maybe also source n. We propose a simple model involving curvature of shock trajectories in magnetic fields through which the origin of the explosion and the centre defined by extrapolated proper motions of BN, I and n may be brought into spatial coincidence.   }
   {}
   

   \maketitle



\section{Introduction}

The Orion Molecular Cloud 1 (\object{OMC-1}) is the nearest and the most studied high mass star-forming region.  The Orion Kleinmann-Low (\object{Orion-KL}) nebula, which harbors an IR luminosity of $\sim$10$\rm^{5}$~L${\sun}$ \citep{Wynn-Williams:1984}, is a region of \object{OMC-1} where two spectacular outflows of different character are found emanating from the same confined region of a few arcseconds \citep[1$\arcsec$ = 414$\pm$7~AU at the distance of Orion,][]{Menten:2007,Kim:2008}. One outflow is a wide-angle high-velocity outflow, at 30~km~s$^{-1}$ to several hundred~km~s$^{-1}$, oriented in a northwest-southeast (NW-SE) direction \citep[][]{Allen:1993,Chernin:1996,Schultz:1999,Odell:2001a,Doi:2002} whereas the second, the so-called low-velocity outflow at 18$\pm$8~km~s$^{-1}$, presents a northeast-southwest (NE-SW) direction projected onto the plane of the sky \citep[]{Genzel:1989,Blake:1996,Stolovy:1998,Greenhill:1998,Nissen:2007,Plambeck:2009,Goddi:2009}.

The energetic outflows \citep{Allen:1993,Odell:2001a} in \object{OMC-1} have prompted many studies seeking their ages and origins. 
The age of the fast outflow is reported to be 500 to 600~years \citep{Doi:2002,Bally:2011,Goddi:2011}. 
\citet{Genzel:1981} reported an age of 3000~years for the 18~km~s$^{-1}$ outflow. However, the recent SiO observations of \citet{Plambeck:2009} carried out with CARMA seem to indicate an age $\ll$~3000~years. 
Nonetheless, our understanding of this region has been transformed by high spatial resolution imaging in molecular hydrogen for which resolution can be better than 100~mas \citep{Lacombe:2004} in the K-band around 2~$\mu$m, in the MIR continuum \citep{Shuping:2004} and through radio-interferometry of a range of species, most recently methyl formate, acetone, acetic acid and dimethyl ether \citep{Favre:2011,Favre:2011a,Friedel:2005,Friedel:2008,Guelin:2008}, methyl cyanide \citep{Wang:2010} and ammonia \citep{Goddi:2011a}, including numerous maser observations in SiO, OH and H$\rm_{2}$O \citep{Doeleman:2002,Goddi:2009,Goddi:2009a,Kim:2008,Matthews:2010,Wright:1995,Johnston:1989,Cohen:2006,Genzel:1977,Genzel:1981,Matveenko:2000,Matveyenko:2007}. Nevertheless the driving sources of the outflows still remain uncertain although it is at least apparent that the high and low velocity outflows are largely independent of each other and arise through different phenomena. 

While it is evident that convulsive events associated with the formation and interaction of embedded high mass stars are the origin of the fast outflow in \object{OMC-1} \citep{Nissen:2007}, the centre of this region presents many candidates \citep{Shuping:2004,Dougados:1993}. A detailed case has recently been made that the origin of the high-velocity outflow results from the disintegration of a non-hierarchical multiple system \citep{Bally:2011} where it has been proposed that a close encounter of I, n and BN, $\sim$500 years in the past, may have triggered the expulsion of material \citep{Bally:2008a,Zapata:2009,Bally:2011}. This model is based upon Very Large Array (VLA) proper motion measurements of the objects BN, I and n~\citep{Gomez:2005,Gomez:2008,Rodriguez:2005}. New work of \citet{Goddi:2011}, again using the VLA but at the higher frequency of 43~GHz, has questioned the involvement of source n while confirming the collision at essentially the same position and a similar epoch between BN and I.
These authors propose that source I formed a hard binary in the course of the encounter with BN. It has also been suggested in a quite different model that an encounter between BN and I, through ejection of BN from the Trapezium cluster, may have powered the outflow \citep{Tan:2004}. A further suggestion is that the nearby radio source SMA1 \citep{Beuther:2008} may be the source of the explosion. In addition, in Submillimeter Array (SMA) observations of the $^{13}$CO (J = 2--1) line \citet{Zapata:2011a} identified a structure suggested to be an expanding bubble with an estimated age of 500--1000~years associated with the same central zone. Suffice it to say that the nature of the encounter which led to the explosive event 500-600~years ago is still a matter of contention but that with a high degree of confidence we know that BN, I and perhaps n suffered a close encounter whose position is known to~$\sim$1$\arcsec$, irrespective of the involvement of source n.    

The low-velocity NE-SW outflow may originate from the radio source I \citep{Menten:1995,Greenhill:2004,Beuther:2006}. This deeply embedded high-mass young stellar object presents a disk \citep{Matthews:2010} perpendicular to the associated NE-SW ground-state SiO maser emission \citep{Goddi:2009,Plambeck:2009}. Alternatively, or in addition, source n may play a significant role since it is also a massive young stellar object with a possible accretion disk \citep{Shuping:2004,Greenhill:2004}. More detail on the nature and geometry of the low velocity outflow region may be found in \citet{Nissen:2007}, based on K-band H$\rm_{2}$ observations. In \citet{Nissen:2007} blue-shifted components are suggested to be the NIR counterpart of the radio outflow. Here we concentrate largely upon the fast outflow introducing the low velocity outflow separately to form a more complete picture of the dynamics.      

 One purpose of the present work is to clarify the geometries of the outflows mentioned above through creation of 3D images of H$_2$ NIR emission. Such images were first described in \citet{Nissen:2008}. Proper motion measurements \citep{Cunningham:2006,Bally:2011} of H$\rm_{2}$ emission at 2.122~$\mu$m with independent data for radial velocities \citep{Gustafsson:2003,Nissen:2007} are combined here to construct 3D motions as in \citet{Nissen:2008}. The bright emission employed to measure the velocity of the disturbance passing through the gas is due very largely to C-type magnetic shocks \citep{Colgan:2007,Kristensen:2008,Kristensen:2007,Kristensen:2003,Takami:2002}. Shock emission is superposed on a nearly ubiquitous background of photon-dominated excitation, typically at the 10$\%$ brightness level. The observations used in this work are briefly reviewed in Sect.~\ref{sec:Observations}. 
 
With a model which initially assumes rectilinear ballistic trajectories, proper motion data \textit{(i)} enable the determination of the approximate location of the expansion centre \citep[already reported in][]{Cunningham:2006,Bally:2011} and \textit{(ii)} give an estimate of the time before present that the explosion took place. The latter is described in Sect.~\ref{sec:modeloutflowbackintime}. In Sect. \ref{sec:modeloutflowevol}, we construct 3D images of the H$\rm_{2}$ emission in \object{OMC-1}. This allows views of the emission from any chosen vantage point and also a reconstruction of the Orion explosion in 3D. 
\citet{Bally:2011} noted a systematic discrepancy between the location of the explosion centre and the point (in projection) of close encounter of BN, I and n, a discrepancy which in the present work, using a subset of the proper motion data and taking into account results in~\citet{Goddi:2011}, appears somewhat heightened. In Sect. \ref{sec:magndefl} we show that an inhomogeneous magnetic field could lead to curved trajectories of ejecta from a central explosion, removing this important deficiency.



\section{Observations: proper motion and radial velocity data for molecular hydrogen, H$\rm_{2}$} 
\label{sec:Observations}
 
\subsection{Proper motion measurements}

Proper motions for the individual features in the \object{OMC-1} outflow were measured using images obtained with the following telescopes: i) the Anglo-Australian Telescope \citep[on 13 September 1992, see][]{Allen:1993}, ii) the Subaru Telescope \citep[on the 11 and 13 January 1999, see][]{Kaifu:2000}, iii) and the Astrophysical Research Consortium 3.5~m telescope at the Apache Point Observatory \citep[on 21~November~2004, see details in][]{Cunningham:2006,Bally:2011}. In all, 194~proper motions were recorded within a 3$\arcmin$ $\times$ 3$\arcmin$ region. 
Proper motions that are used here are in the range of 20~km~s$\rm^{-1}$ up to 120~km~s$\rm^{-1}$ with an uncertainty of  10~km~s$\rm^{-1}$. Discussion of errors in proper motion data may be found in section \ref{comb-data}.

\begin{figure}[h!]
\centering
\renewcommand{\footnoterule}{} 
\includegraphics[width=9cm]{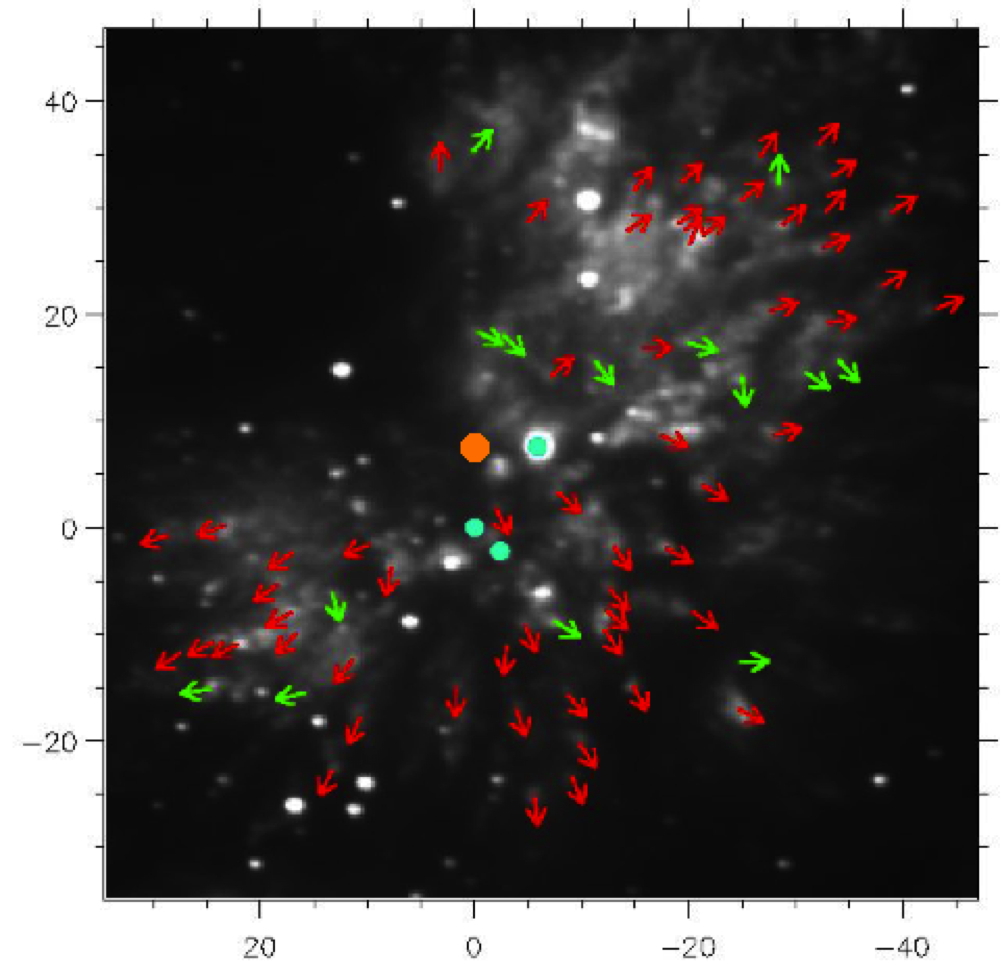}
 \caption{Field of view covering the inner 0.1~pc (1$\arcmin$ $\times$ 1$\arcmin$) of \object{OMC-1} in the present work. The grey-scale shows the emission line v=1--0 S(1) of H$\rm_{2}$ at 2.121~$\mu$m \citep[observations carried out with the 3.6~m CFHT telescope using the GriF instrument, see][]{Gustafsson:2003,Gustafsson:2006,Nissen:2007}. Green and red arrows, representative of the 71 data for which there exist both proper motions and radial velocities, show the measured position angles. Those arrows which point with a deviation of less than 30$\degr$ from a computed centre of expansion based on all 194 proper motion data \citep{Cunningham:2006,Bally:2011} (the orange dot -- see text) are shown in red and those that show a deviation $>$~30$\degr$ are in green. Positions of the BN object ($\alpha_{J2000}$ = 05$^{h}$35$^{m}$14$\fs$1094, $\delta_{J2000}$ = -05$\degr$22$\arcmin$22$\farcs$724), the source n ($\alpha_{J2000}$ = 05$^{h}$35$^{m}$14$\fs$3571, $\delta_{J2000}$ = -05$\degr$22$\arcmin$32$\farcs$719) and the radio source I  ($\alpha_{J2000}$ = 05$^{h}$35$^{m}$14$\fs$5141, $\delta_{J2000}$ = -05$\degr$22$\arcmin$30$\farcs$575) \citep{Goddi:2011} -- turquoise dots -- are indicated. The coordinates of (0,0) are those of the radio source I. Units are arcseconds, where 1$\arcsec$ = 414$\pm7$~AU \citep{Menten:2007,Kim:2008}.
 }
             \label{Figure1}%
    \end{figure}

The directions but not the magnitudes of these proper motions within the 1$\arcmin$ $\times$ 1$\arcmin$ region covered here are shown in Fig. \ref{Figure1}. On the basis of these proper motion data, \citet{Cunningham:2006} and \citet{Bally:2011} using rectilinear trajectories showed that the geometrical centre of expansion of the fast outflow lies at $\alpha_{J2000}$ = 05$^{h}$35$^{m}$14$\fs$50, $\delta_{J2000}$ = -05$\degr$22$\arcmin$23$\farcs$00. 
In Fig. \ref{Figure1}, red arrows indicate features with position angles which point within 30$\degr$ of the centre of expansion. The latter is shown as a deep blue dot and is  computed from all 194 proper motion data.
%
 
\subsection{Radial velocity measurements}

\begin{figure}[h!]
\centering
\renewcommand{\footnoterule}{}  
\includegraphics[width=8.8cm]{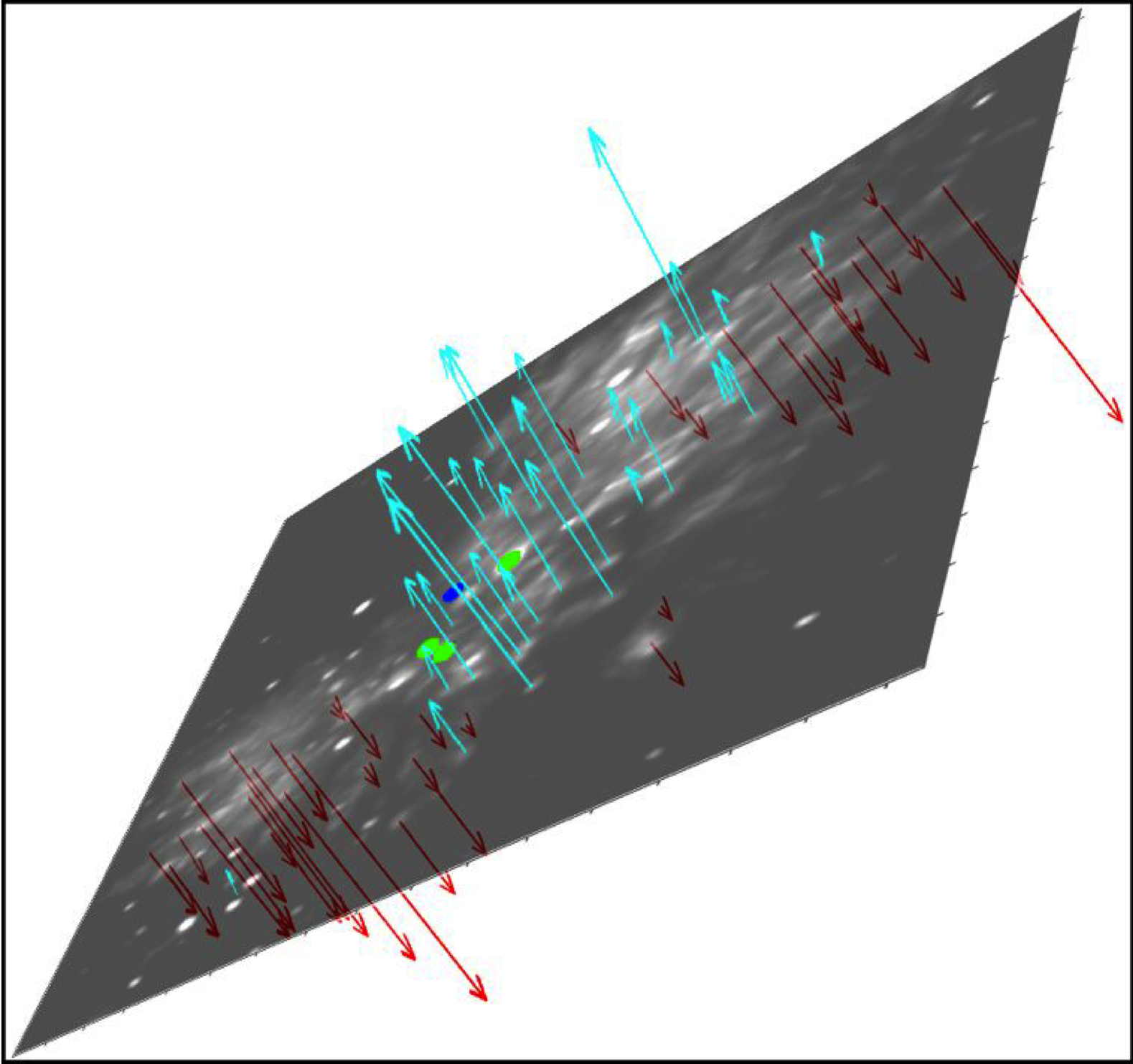}
 \caption{Radial velocities of 71 individual features in the \object{OMC-1} outflow for the same 1$\arcmin$ $\times$ 1$\arcmin$ field of view as in Fig.~\ref{Figure1}. The orientation of the image -- with the Earth to the top left -- illustrates the radial velocities associated with each of the objects for which proper motions are shown in Fig. \ref{Figure1}. Blue and red-shifted emissions are shown as blue and red arrows, respectively. The length of the longest arrows corresponds to 50~km~s$^{-1}$ in the local standard of rest.
 }
             \label{Figure2}%
    \end{figure}

The field of view of radial motions covers the inner 0.1~pc (1$\arcmin$ $\times$ 1$\arcmin$) of \object{OMC-1} and is presented in Fig.~\ref{Figure1}. Radial motion data, ranging up to 50~km~s$^{-1}$ (v$\rm_{LSR}$), were recorded on December 5$\rm^{th}$ 2000 at the 3.6~m Canada-France-Hawaii Telescope (CFHT) using the Fabry-Perot GriF instrument \citep{Clenet:2002} to perform high resolution IR spectroscopy and obtain radial velocities of the H$\rm_{2}$ emission pixel by pixel from the observed Doppler shifts. A detailed description of data and analysis is given in \citet{Gustafsson:2003,Gustafsson:2006} and \citet{Nissen:2007}.
The spatial resolution of these data is 0$\farcs$18. Radial velocities are shown in Fig. \ref{Figure2}.

Seventy-one objects could be unequivocally identified as common to the proper motion and radial data sets. 
Turning to the radial data in Fig.~\ref{Figure2}, the velocities (lsr) cover the range $-$35~km~s$^{-1}$ to $+$50~km~s$^{-1}$ with an average of $\sim$20~km~s$^{-1}$.
The coordinates of the 88 features observed in the GriF data  were converted and then compared to the H$\rm_{2}$ structures in order to obtain the same features within the same area. This reduces the number of objects to 71. Note that of the 17 features excluded, 14 were excluded on the basis that the H$\rm_{2}$ emission was too weak to determine an accurate radial velocity and 3 on the basis that the objects were not sufficiently resolved that a single radial velocity could be assigned.
Further details of the method used for the identification of objects common to both radial and proper motion datasets can be found  in \citet{Nissen:2008}.
%

\subsection{The combined dataset}
\label{comb-data}

It is evident from the position angles shown in Fig.~\ref{Figure1} that some features cannot reasonably be associated with the outflow. Any datum for which the position angle points with a deviation of more than 30$\degr$ from the computed H$\rm_{2}$ origin, that is, any marked with a green arrow in Fig.~\ref{Figure1}, was not included in our analysis. Thus 14 proper motion measurements were excluded by this criterion, leaving 57 features in the dataset. We mention in Sect. 3.2 that this pre-selection has no bearing on our conclusions. The choice of 30$\degr$ arises from a natural dip in the distribution. Indeed seven of the discrepant features form a concerted group which traverses the field 15$\arcsec$-20$\arcsec$ northwest of the central region. This may indicate that a collimated outflow, unrelated to the main \object{OMC-1} flows studied here, crosses the field of view.

Errors \citep[1$\sigma$, see][]{Gustafsson:2003,Cunningham:2006,Nissen:2007} in radial velocities and proper motions are 1 to 2~km~s$^{-1}$ and 10~km~s$^{-1}$, respectively. 
In this connection errors in the proper motions are recorded in \citet{Cunningham:2006} as 20~km~s$^{-1}$ with a mean value of 15~km~s$^{-1}$ in \citet{Bally:2011}. In our data for proper motions we use a subset of those in \citet{Bally:2011} and record an error of 10~km~s$^{-1}$ (1$\sigma$). This lower value arises because the subset of data which we use excludes data (see above) and these exclusions contain a number of less accurate values. 
We also note that sufficiently fast shocks will dissociate H$\rm_{2}$. For pre-shock gas of density 10$^{5}$ -- 10$^{6}$~cm$^{-3}$ and depending on the local magnetic field strength, this places an upper limit of ~$\sim$70~km~s$^{-1}$ on the maximum velocities of features visible in our data \citep{Kristensen:2008}. However, some H$\rm_{2}$ features with proper motions exceeding this limit are observed toward \object{OMC-1}. These features are included in the present study. It would seem likely that these faster objects impact on material which is already in recessional motion, a feature for which there is some observational evidence~\citep{Nissen:2007}. 



\section{The \object{OMC-1} outflow : backwards in time}
\label{sec:modeloutflowbackintime}

In this section proper motions of our 57 chosen objects are traced backwards in time.
Note that here and in all subsequent sections up to Sect. \ref{sec:magndefl} rectilinear ballistic motion is assumed. Proper motion measurements record the passage of a disturbance, that is, of bullets through the surrounding gas. 
We consider first if it is a good approximation to consider this motion as ballistic. We treat the inner bullets shown in Fig.~\ref{Figure1} as slower moving versions of the fast bullets in the so-called \textquotedblleft fingers\textquotedblright \ \citep{Allen:1993} and assume a bullet mass of 5 $\times$ 10$^{-4}$~M${\sun}$ \citep{Burton:1997,Chrysostomou:1997}. We use as an example a very well characterised shock, 20.7$\arcsec$ west and 6.8$\arcsec$ south of the reference star TCC0016, which has been modeled in 3D \citep{Gustafsson:2010} and imaged at 35~AU resolution \citep{Lacombe:2004}. The bullet which gives rise to this emission has a velocity of 50~km~s$^{-1}$ \citep{Gustafsson:2010}, typical of bullets in the inner region, and thus an energy of $\sim$1.25 $\times$ 10$^{36}$~J. The average emission from this region, lozenge shaped and of dimension 0.1$\arcsec$ radius and 0.5$\arcsec$ in length is $\sim$10$^{-5}$~Wm$^{-2}$sr$^{-1}$ in the S(1) v=1-0 H$\rm_{2}$ rovibrational emission line \citep[see Fig. 11 of][]{Nissen:2007}. Given these figures, the total energy emitted in the S(1) line is $\sim$3.3 $\times$ 10$^{33}$~J if the emission continues for $\sim$700 years as our simulations suggest. Emission in the S(1) line makes up $\sim$5-10$\%$ of total H$\rm_{2}$ emission giving an emitted energy of $\sim$4 $\times$ 10$^{34}$~J. 
Allowing for some additional emission lines of other molecules this suggests that $\sim$4$\%$ of the initial energy is lost through radiation, translating into 2$\%$ in the velocity.
Thus disturbances in the gas involve objects which in general have a considerably greater initial kinetic energy than that which is given up in the shocks which they generate, justifying the use of a ballistic model as an acceptable approximation.
 
\subsection{Description of the main \object{OMC-1} outflow}
\label{subsec:desc}

\begin{figure}[h!]
\centering
\includegraphics[width=6.9cm,angle=270]{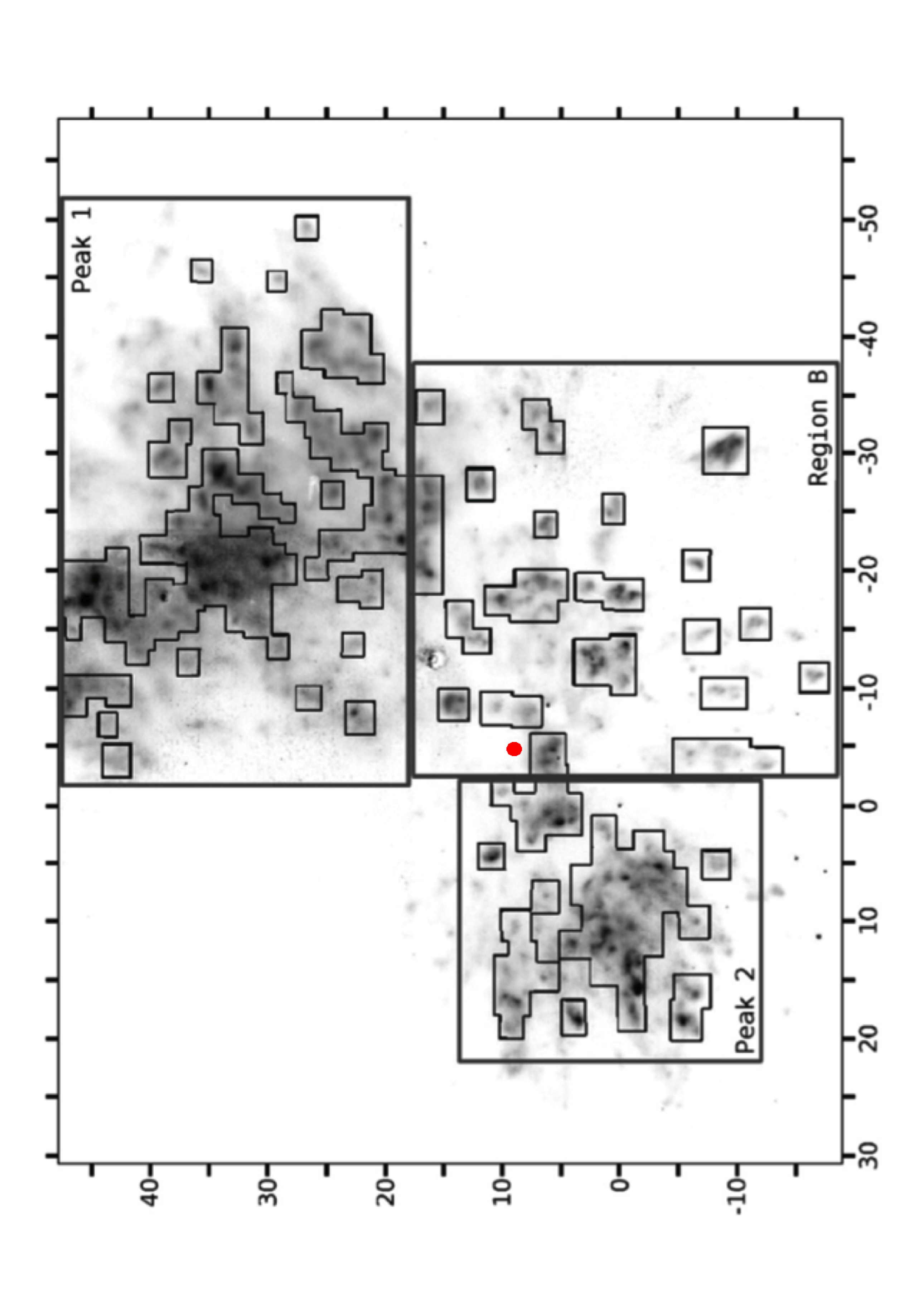}
 \caption{The naming conventions for Peak 1, Peak 2 \citep{Beckwith:1978} and Region B towards \object{OMC-1} superposed on the image of the H$\rm_{2}$  v=1-0 S(1) line at 2.121~$\mu$m. The coordinates of (0,0) are those of the star TCC0016 ($\alpha_{J2000}$ = 05$^{h}$35$^{m}$14$\fs$909, $\delta_{J2000}$ = -05$\degr$22$\arcmin$39$\farcs$31), \citet{Vannier:2001}. The red dot indicates the position of source I. Units are arcseconds, where 1$\arcsec$ = 414$\pm7$~AU \citep{Menten:2007,Kim:2008}. Figure taken from \citet{Nissen:2007}.}
             \label{Fig.Refregions}%
    \end{figure}

Constituents of the fast and slow \object{OMC-1} outflows exhibited by H$\rm_{2}$ data within 1$\arcmin$ $\times$ 1$\arcmin$ are shown in Fig. \ref{Fig.Refregions}. The fast outflow is represented by two regions, traditionally called Peaks 1 and 2~\citep{Beckwith:1978} and labelled as such in Fig. \ref{Fig.Refregions}. The slow outflow, first detected in H$\rm_{2}$ in~\citet{Gustafsson:2003}, and which we call Region B (see Fig. \ref{Fig.Refregions}), is composed entirely of blue-shifted shock zones \citep{Nissen:2007}.

\subsection{The time of the explosion}

\begin{figure}[h!]
\centering
\includegraphics[width=9.9cm,angle=0]{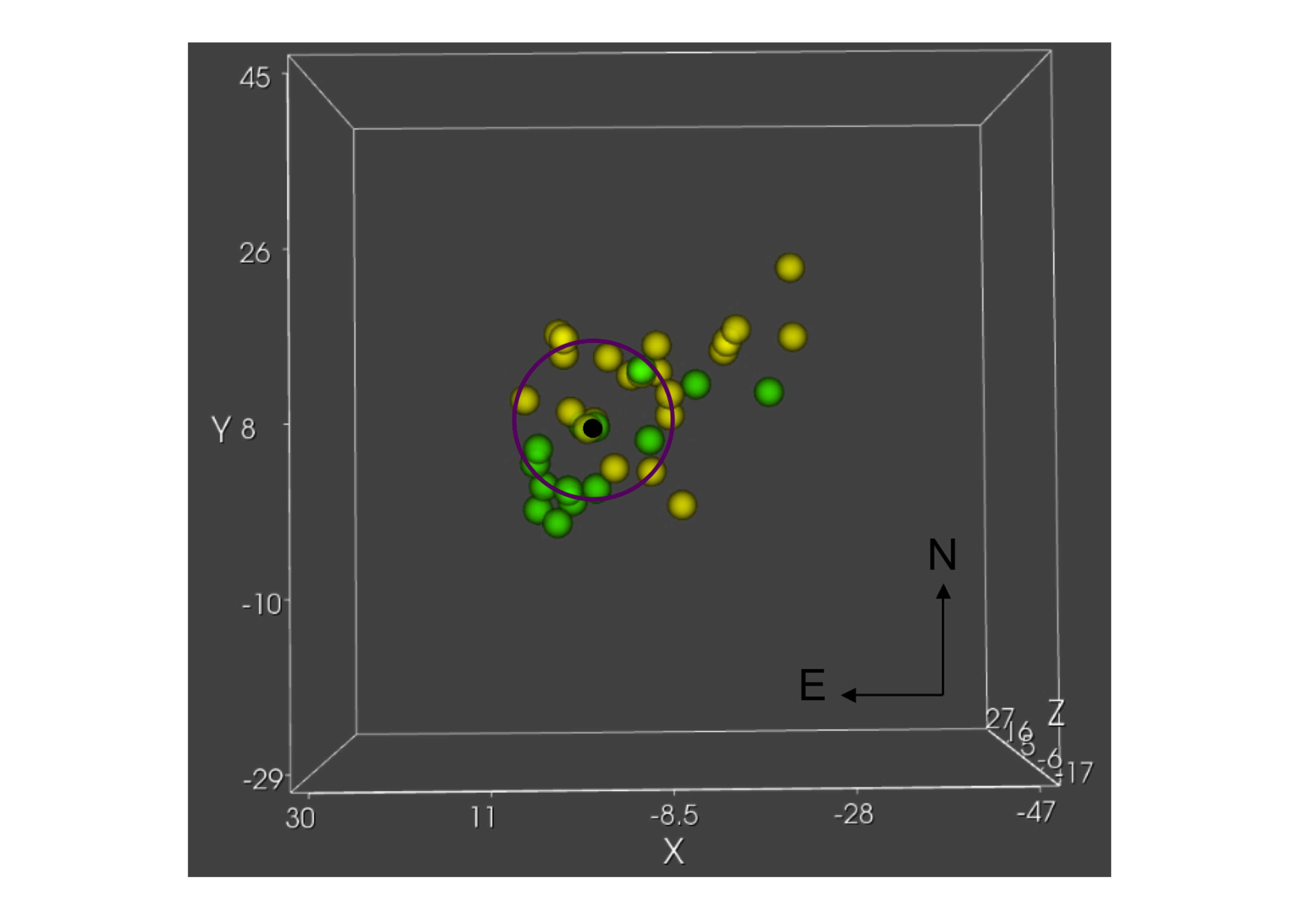}
 \caption{Disposition of the Peak 1 and Peak 2 objects in yellow and green, respectively at the epoch time zero of the explosion seen in the line-of-sight (i.e. perpendicular to the plane of the sky xy). Thirty-six features are displayed in this figure. The dark dot is the computed expansion centre~\citep{Nissen:2008,Bally:2011}. The ring-like structure which forms when proper motions are run backwards in time is highlighted in violet. This structure is the starting point for the 3D-simulations illustrated in subsequent figures. Units are in arcseconds.}
             \label{Fig.Ringt0}%
    \end{figure}

Proper motions for the subset of 36 objects in Peaks 1 and 2 were run backwards in time, assuming linear ballistic motion. Rather than converging on a single spot, these motions converged to a ring-like structure shown in Fig. \ref{Fig.Ringt0} at 720$\pm$25~years in the past. This time is an upper bound since deceleration has of course been ignored in a ballistic model. The computed centre of expansion in \citet{Bally:2011} lies at $\alpha_{J2000}$ = 05$\rm^{h}$35$\rm^{m}$14$\fs$50, $\delta_{J2000}$ = -05$\degr$22$\arcmin$23$\farcs$00) and is localized about 7.5$\arcsec$ to the north of the radio source I \citep[J2000,][]{Plambeck:1995,Goddi:2011} and 6$\arcsec$ to the east of the BN object \citep[J2000,][]{Menten:1995,Goddi:2011}. Shown as a deep blue dot in Fig.~\ref{Fig.Ringt0}, this is found to be located at the centre of the ring illustrated in that figure.  

The observed ejecta appear to originate from two caps of material, one associated with Peak 1 and the other with Peak 2, as shown in yellow and green in Fig. \ref{Fig.Ringt0}, respectively, with some obvious outliers. There are 5 bullets in the centre, three from Peak 1 and two from Peak 2, which as we have seen is coincident within measurement errors with the computed centre of expansion, based on all 194 proper motion data. If motions of the objects are allowed to develop further into the past the bullets do not become more confined. Rather, the two caps pass through each other and the ring structure is lost within 25 years. These characteristics led us to consider the ring structure in Fig.~\ref{Fig.Ringt0} as the structure corresponding to the epoch of the explosion.   

The observed ring-structure has a diameter of about 6000~AU ($\sim$15$\arcsec$). The empty lane between the centre and rim seen in Fig. \ref{Fig.Ringt0} shows that the structure is not a sphere given that we are observing clumps of dense shocked gas. The structure is robust to inclusion of errors and may be clearly discerned with use of all 194 proper motion data without introduction of any of the selection criteria adopted here. Results also show that the ring structure is largely populated from objects within $\sim$0.1~pc of the expansion centre. Thus we do not propose this structure as the precursor of the very fast moving bullets which make up the outer fingers of emission \citep{Allen:1993}. Figure 2 of \citet{Bally:2011} illustrates these points.  

Further in connection with the ring, we note that the ring is well described as circular and this suggests it is not tilted in the plane of the sky. Moreover it forms by a remarkable coincidence of individual and numerically very disparate values of proper and radial motions. An apparent lack of structure up to $\sim$650 years in the past leads to a rather abrupt appearance of the ring structure ($\pm$ $\le$25-50 years) running further backwards in time. Beyond 720 years in the past the ring structure rapidly disappears.  

The apparent origin of so many bullets in a well defined structure near the computed centre at one time lends considerable weight to a single explosion hypothesis to describe the inner part of the fast outflow lobes. However there are aspects of the data that highlight the limitation of our present simple interpretation. For example, three Peak 2 bullets in Fig. \ref{Fig.Ringt0} appear to originate from Peak 1. In addition, as noted above, the explosion centre in Figs. \ref{Figure1} and \ref{Fig.Ringt0} differs from the position of closest encounter of I, BN (and n)~\citep{Bally:2011,Goddi:2011,Zapata:2011a}, a point which we discuss in more detail in Sect. \ref{sec:magndefl}. 

At all events our analysis indicates that the ejection occurred $\le$ 720~years ago, which is in reasonable agreement with the range of dates of 500-560 years deduced in previous studies \citep{Bally:2011,Goddi:2011,Gomez:2008,Gomez:2005,Rodriguez:2005,Zapata:2009} and data show, as in \citet{Bally:2011}, that the close encounter of BN, I and n is a feasible explanation for the Orion explosion.



\section{A 3D view of the \object{OMC-1} outflow}
\label{sec:modeloutflowevol}

Our purpose here is to combine proper motion and radial data to create a 3D model of the Orion explosion as observed in H$\rm_{2}$ emission. We remark first that proper motion measurements record the passage of a shock through the surrounding gas, where this disturbance, marked by H$\rm_{2}$ excitation, is generated through material - bullets - flung out at high velocity due to an explosive event. Radial velocities obtained from GriF data however record the actual motion of the H$_2$ itself and not directly of the bullets. We base all 3D constructions that follow on the assumption that the velocity in the radial direction can be treated as a measure of the radial velocity of the disturbance and thus that the gas that is emitting has been accelerated to the full shock velocity. In fact C-type shock models show that the peak emission in the gas is reached when the gas is moving typically at a velocity which is 5~km~s$^{-1}$ or more below the velocity of the shock and thus the radial motion of the hot H$\rm_2$ is not a precise measure of the radial component of the motion of a bullet. This velocity difference arises because it is the conversion of kinetic energy into internal energy which generates the hot H$\rm_2$. However since motion is concentrated in the plane of the sky the associated systematic error in the estimate of 3D velocities is $<$2-3~km~s$^{-1}$ in values which may be typically 50-60~km~s$^{-1}$~\citep{Kristensen:2008}. This small systematic effect has been ignored here.


\subsection{The fast outflow: Peaks 1 and 2}
\label{sec:fast}

Based on the above conclusion that the ejecta originate from a
ring-like structure, and not some more spherical distribution, 
the structure shown in Fig. \ref{Fig.Ringt0} is taken to mark the epoch of the explosion and all bullets that form Peaks 1 and 2 are assumed to lie in the plane of the sky at this epoch. Motion now runs forward in time, including both radial and proper velocities, evolving towards the present and tracing out the formation of the lobes of the explosion. The 3D structure of the H$\rm_{2}$ bullets 720~years later is shown in Fig.~\ref{Fig.3dstructf}. Snapshots at 40 year intervals, including slow outflow objects, are shown in Appendix \ref{Appendix-movie} (Fig. \ref{3dmovreco})\footnote{Animations allowing a visualization from this and other vantage points can be obtained upon request to the authors.}.

\begin{figure}[h!]
\centering
\includegraphics[width=9cm,angle=0]{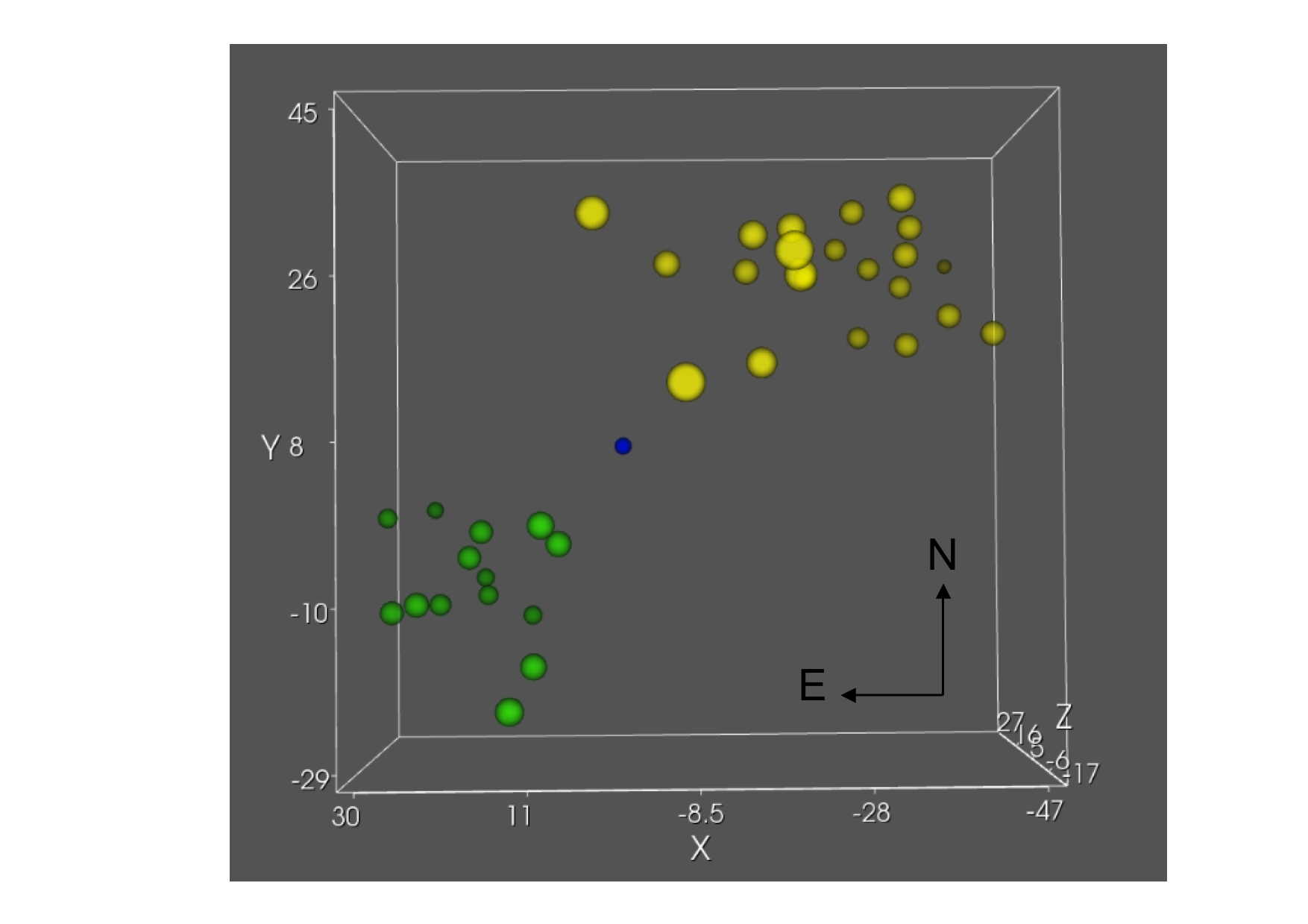}
 \caption{Disposition of Peak 1 and Peak 2 objects in yellow and green respectively at the present epoch seen in the line-of-sight. Thirty-six features are displayed in this figure. This image is a reconstruction from Fig. \ref{Fig.Ringt0} after 720~years. The dark blue dot  is the computed expansion centre, as earlier. Larger and brighter objects are closer. Units are in arcseconds.}
             \label{Fig.3dstructf}%
    \end{figure}

\begin{figure}[h!]
\centering
\includegraphics[width=9cm,angle=0]{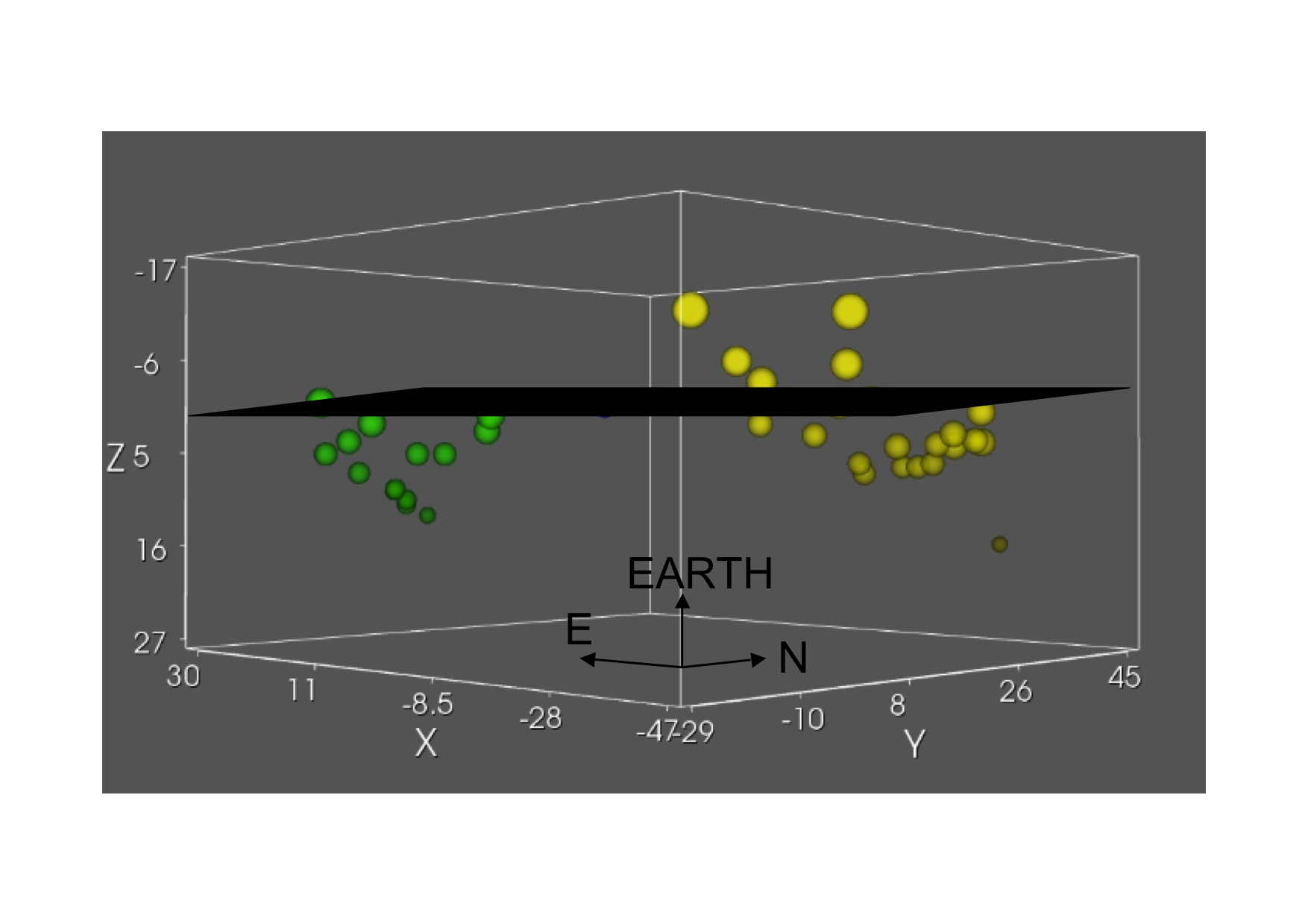}
 \caption{Disposition of Peak 1 and Peak 2 objects in yellow and green respectively at the present epoch viewed from an angle of 45$\degr$ beneath the Galactic plane, that is, the xz plane. Earth lies in a direction vertically upwards. The plane of the sky is shown corresponding to z = 0, passing through the computed expansion centre. An angle of $\sim$150$\degr$ between the axes of the Peak 1 and 2 flows may be seen. Units are in arcseconds.}
             \label{Fig.angleP1P2}%
    \end{figure}

The appearance of Peak 1 and 2 bullets at a viewing angle of 45$\degr$ from below the plane of the Galaxy is illustrated in Fig. \ref{Fig.angleP1P2}. This figure shows that Peaks 1 and 2 regions are each bent away from the Earth at an angle of $\sim$15$\degr$ out of the plane of the sky. The two lobes can be seen to lie at an estimated value of~$\sim$150$\degr_{-20\degr}^{+10\degr}$ rather than opposed at 180$\degr$. The true angle may be smaller than 150$\degr$ because of the observational selection of proper motions and radial velocities. Since both flows are angled away from the Earth, the effect does not arise through dust obscuration. Note that the effect is already apparent in the radial data shown in Fig.~\ref{Figure2}. The presence of an inhomogeneous magnetic field could be responsible for this phenomenon, as discussed in Sect.~\ref{sec:magndefl}.

\subsection{The slow outflow: Region B}

Region B is a continuous or punctuated outflow running from the northeast (NE) to southwest (SW), located southwest of radio sources I and n (see section \ref{subsec:desc} and Fig. \ref{Fig.Refregions}). The reason for the choice of a continuous outflow for Region B rather than an origin in a single explosive event is given below: at all events the nature of the Region B outflow is quite distinct from the explosive fast outflow to which it lies roughly at right-angles in the plane of the sky. There is clear observational evidence, set out in \citet{Nissen:2007}, that the Region B outflow originates from a central zone close to or coincident with source I. However it remains uncertain whether the origin of the outflow is source I, source n or some other unknown object.  Thus, \citet{Nissen:2007} have suggested that Region B is the IR counterpart of a continuous outflow detected in the radio from either source I or n \citep{Greenhill:1998,Greenhill:2004,Matthews:2010,Goddi:2009,Plambeck:2009,Wright:1996,Genzel:1981}. Recent v=0 SiO maser data \citep{Goddi:2009,Plambeck:2009} show very clearly the association of this extended SiO emission with source I. The northeast--southwest orientation of the SiO emission provides further evidence that the Region B outflow is associated with source I on the geometrical grounds of the flow orientation. 
In conclusion, since there is good evidence in \citet{Nissen:2007} and from more recent SiO masers observations, we feel justified in treating Region B as a counterpart of a continuous (or punctuated) outflow arising from the central zone close to the origin or at the origin of the explosion which gave rise to Peaks 1 and 2. 

The difficulty accompanying this characterization of Region B is that the BN, I (n) encounter took place $\sim$700 years ago, using the figure derived from our data. The accompanying violence which gave rise to Peaks 1 and 2 at that time would disrupt any outflow from I itself or from the near vicinity of I, if only since outflow B is of much lower total energy content. Thus we require that the outflow B is no older than $\sim$700 years and it would seem necessary to put on one side the age estimate of \citet{Genzel:1981} of 3000 years and rather adopt the value of a few hundred years consistent for example with data in \citet{Plambeck:2009}. This is itself consistent with our finding that extrapolation of ballistic motion of objects in Region B backwards in time for the full 720~years (see section 4.1) results in some objects in the Region B outflow passing through the central zone and retreating out the other side. Our inference is that in these cases the H$\rm_{2}$ emission traces more recently emitted objects in a continuous outflow from the central zone. Using this model, data for Region B are now combined with those for the fast outflows already illustrated.

\subsection{Peaks 1, 2 and Region B objects}

We now seek to combine all objects in our sample. We first explain how Region B objects can be incorporated, given the considerations in the preceding section. As noted, the Region B outflow is of quite different character to that of the Peaks 1 and 2. The latter originate from a single explosive event whereas Region B objects constitute members of a much more familiar form of continuous or punctuated outflow associated with star formation. The Region B outflow is therefore treated as emanating from a single zone with continuous replenishment as time passes. If we take the present positions of Region B objects and reverse them in time, such a reversal would therefore endure for a length of time which is different for every object. This set of times is given by the times required for objects to find their way back to the zone of origin of the outflow assuming rectilinear, ballistic motion. In our simulation the zone of origin is identified as any point on a line bisecting Peaks 1 and 2. Thus if we now move forwards in time, tracing the spatiotemporal evolution of the Region B outflow, new objects are created at this set of times, lying between 720~years in the past and up to close to the present.
Region B objects as a whole involve objects with radial velocities in the range -24~km~s~$^{-1}$ to -8~km~s~$^{-1}$ and proper motions from 28~km~s~$^{-1}$ to 102~km~s~$^{-1}$. There is a notable difference between these velocities and values  for 22~GHz H$\rm_{2}$O masers which are clustered around 18~km~s$^{-1}$ \citep{Genzel:1981}. H$\rm_{2}$O masers form where the temperature is very much lower and the density very much greater than in the region where H$\rm_{2}$ emission is observed \citep{Yates:1997}. Hence, we expect H$\rm_{2}$O masers and H$\rm_{2}$ features to trace different portions of the flow.
 %
\begin{figure*}
\centering
\includegraphics[width=19cm,angle=0]{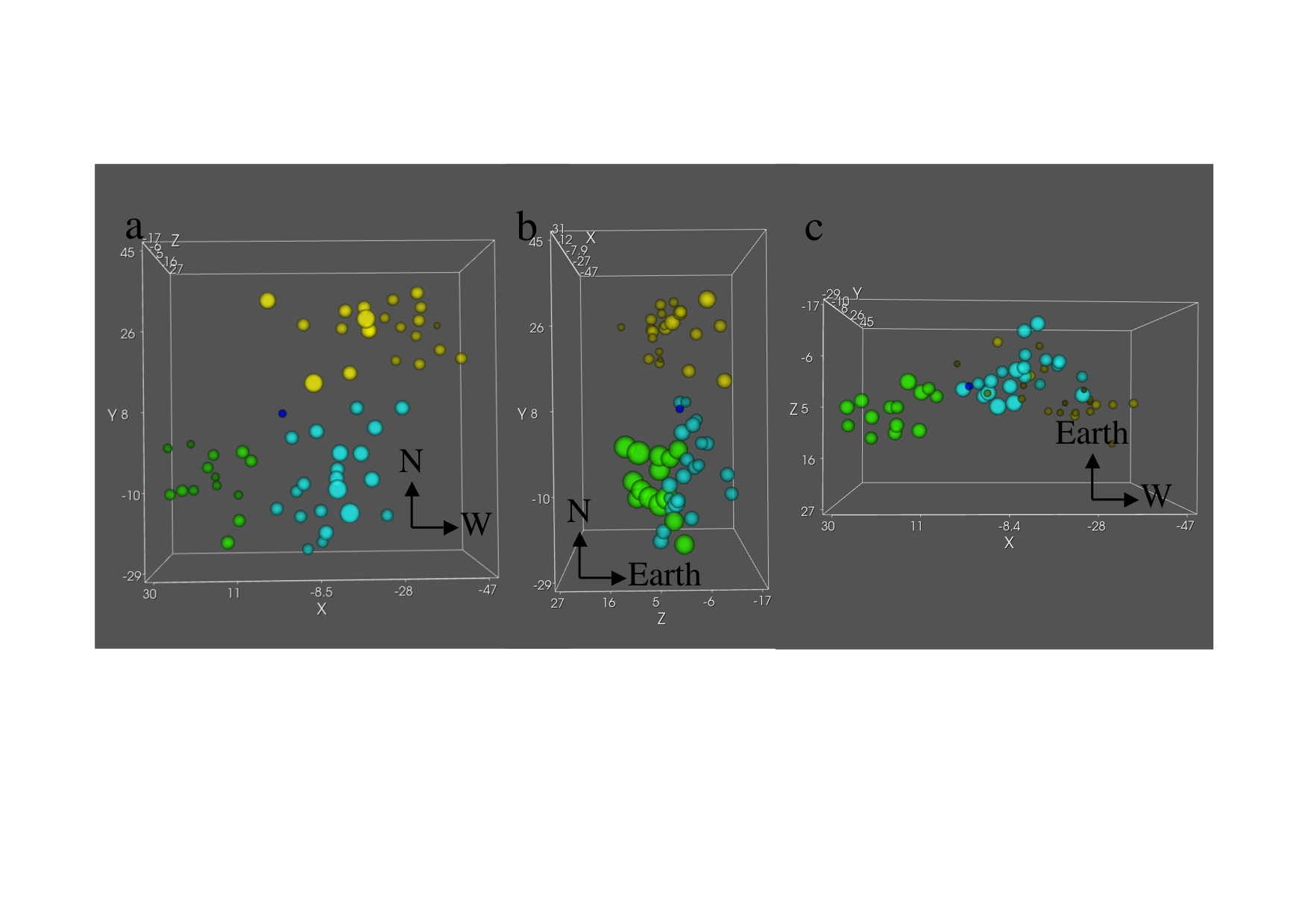}
 \caption{Peak 1, Peak 2 and Region B objects at the present day in yellow, green and turquoise, respectively. The deep blue dot is the computed expansion centre. Units are in arcseconds. a) The view as seen in the line-of-sight, b) the view as seen from the Sagittarius arm of the Galaxy (normal to the yz-plane), c) the view as seen from beneath the Galactic plane (xz).}
             \label{Fig.regionBP1P2}%
    \end{figure*}

The present day appearance of the objects associated with Peak 1, Peak 2 and Region B are shown in Fig. \ref{Fig.regionBP1P2} from different vantage points: a) as seen from Earth, b) as seen from the direction of the Sagittarius arm of the Galaxy\footnote{Orion lies in the Galactic plane and the Sagittarius arm is at right angles to the Earth -- \object{OMC-1} direction.} and c) as seen from beneath the Galactic plane.
More specifically, Fig. \ref{Fig.regionBP1P2}c and to some extent 7a, show how Region B protrudes towards the Earth (along the negative z axis) and points towards the west. Fig. \ref{Fig.regionBP1P2}b also shows that Region B is angled strongly south, noting that this effect may be exaggerated through the omission of blue-shifted objects with proper motions less than 10~km~s$^{-1}$. 

Thus Region B represents a flow in a south-westerly direction which is angled to some small extent towards the Earth. This geometry can be rationalized if source I is the source of the Region B outflow, as posited in \citet{Nissen:2007}. Consider the plane defined by the direction of proper motion of source I and the normal to the sky. According to results in \citet{Goddi:2011} this is a plane tilted 20$\degr$ south to the horizontal with the motion of I pointing south-east. Source I is recorded in \citet{Goddi:2011} as moving at 11.5~km~s$^{-1}$ which is reported to be the full velocity of source I with respect to the nebula. Source I is thus moving in the plane of the sky. An outflow from I will therefore appear to be projected in a manner which combines the motion of I and the radial outflow velocity of Region  B, whose average is 18~km~s$^{-1}$, a figure about which values are quite tightly clustered \citep[see Fig. 5 of][]{Nissen:2007}. 
This will bend the flow to the south and make it a little less westerly compared with the orientation of the outflow in the moving frame of source I. The same consideration would apply to outflows associated with extended SiO maser emission in v=0 \citep{Goddi:2009,Plambeck:2009,Goddi:2009a,Cho:2005} as alluded to in \citet{Plambeck:2009}. In this connection the orientation of the Region B outflow, as seen most clearly in Fig. \ref{Fig.regionBP1P2}c, corresponds well with the axis of the SiO v=0 outflow recorded in \citet{Goddi:2009} as mentioned in section 4.2. In both cases these show a position angle of $\sim$230$\degr$ measured north to east from the axis defined by the fast outflow.
The direction of the Region B outflow would also appear to be at right-angles to the orientation of an edge-on disk around I, whose presence is inferred from SiO v=1,2 maser observations in \citet{Matthews:2010} as mentioned in the introduction. 
The Region B outflow will be sprayed out to some extent due to dispersion about values of the radial velocity and proper motion velocities. Objects will therefore be disposed relative to the source of the outflow depending both on the time of ejection of the bullets and their individual proper and radial motions.
The protrusion of any Region B bullet out of the plane of the sky and towards the Earth will be tan$\rm^{-1}$(radial velocity/proper motion velocity). This is $\sim$20$\degr$ using a typical value of radial velocity of 18~km~s$^{-1}$ and proper motion velocity of 50~km~s$^{-1}$. 

Above for simplicity we have arbitrarily used the expansion centre for the origin of the Region B outflow. If however we were to use source I, say, then features observed in Region B would have to be introduced slightly later into the model. This is due to the geometrical effect of the shift of the SE-NW line bisecting Peaks 1 and Peaks 2 that now runs through source I instead of the centre of expansion. Accompanying changes in the positions of Region B objects in the 3D images do not exceed 3$\arcsec$.


\section{Magnetic deflection of the ejecta in the \object{OMC-1} explosion}
\label{sec:magndefl}

A detailed and convincing model has been developed \citep{Bally:2011,Goddi:2011,Gomez:2008,Gomez:2005,Rodriguez:2005} in which the fast outflow in Orion was caused by the close encounter of sources BN, I and n $\sim$500~years in the past, notwithstanding the doubts recently raised about the involvement of source n. A potential problem with the hypothesis, noted in \citet{Bally:2011} and to which we have already referred, is that the proper motions of the 194 ejecta recorded in that paper, produced as rectilinear, intersect at a centre which is displaced from the known position of the explosion by $\sim$5$\arcsec$ to the northeast. Since the point of intersection has itself an uncertainty of 5$\arcsec$, \citet{Bally:2011} concluded that the position of the explosion and of the intersection were not different at a level to cause too much concern. However this remains a nagging problem and it is heightened in the present work where the centre of the ring in Fig. \ref{Fig.Ringt0} is 5.5$\arcsec$~south and 2.5$\arcsec$~west, that is, 6$\arcsec$ displaced from the BN, I, n centre of encounter or 7$\arcsec$ displaced relative to the centre presented in \citet{Goddi:2011}. In addition, the present apparent ejecta centre in Fig. \ref{Fig.Ringt0} is rather closely spatially confined by 5 bullets occupying a region of diameter $<$3$\arcsec$ with the BN, I, (n) centre of encounter lying on or very near the rim of the circular structure in Fig. \ref{Fig.Ringt0}. Errors in the absolute position of the BN, I, (n) encounter are small by comparison \citep[$\sim$1$\arcsec$,][]{Goddi:2011} and do little to alleviate this problem.  

A further issue is that the outflows are unexpectedly bent away from us in the plane of the sky in both Peaks 1 and 2, as recorded in section \ref{sec:modeloutflowevol}.1 and figure \ref{Fig.angleP1P2}. Such a geometry does not arise naturally from the extensive models in \citet{Bally:2011} or \citet{Goddi:2011}. Both these difficulties may be simultaneously resolved if the ejecta have been moving since the time of the explosion in an inhomogeneous magnetic field showing an organised structure on the scale of the inner 0.1 pc of the outflow. As we have noted, shocks and the accompanying vibrationally excited H$\rm_{2}$ are generated through the rapid passage through the medium of masses of material -- the bullets. The physical nature of these bullets is unknown but it is reasonable to assume that by whatever of the three explosion mechanisms proposed for example in \citet{Bally:2011}, the material of the bullets may have achieved substantial ionization. The trajectories of these ejecta will then become curved in the presence of a magnetic field gradient. At this juncture we should add that it is possible that the deflection is purely hydrodynamic. Specifically, the dense \object{OMC-1} dust ridge runs mainly north-south. Thus outflow components moving north-west may run into a density gradient which naturally will tend to deflect the ejecta towards the west. We do not pursue this line of argument further, if only through the lack of observational constraints.     

To return to the magnetic model, deflection of a plasma jet in a magnetic field gradient is a well-known phenomenon in plasma physics and engineering \citep[e.g.][]{Demichev:1965,Timoshenko:2008}. The phenomenon does not appear to have been much considered in the astrophysical context in contrast to magnetic collimation or magnetic centrifugal acceleration \citep[e.g.][]{Bogovalov:1999,Banerjee:2006}. The most simple manner to appreciate the phenomenon is through the magnetic pressure, $\rm B^{2}$/2$\rm \mu\rm_{0}$ \citep{Demichev:1965}. If this increases radially, in the direction of the radius vector $\vec{r}$, then the excess pressure on the opposite or outer side of any one of the ejecta will be greater than on the inner side, giving rise to motion in a circular path. This approach glosses over the physics involving the detailed motions of the electrons and ions but allows an estimate to be made of the radius of curvature, r, of the path of plasma ejecta. The relatively small effect of deflection in the direction of $\vec{r} \times \vec{B}$, that is, perpendicular to the radius vector, is ignored here \citep{Timoshenko:2008}.

\begin{figure*}
\centering
\includegraphics[width=15cm,angle=0]{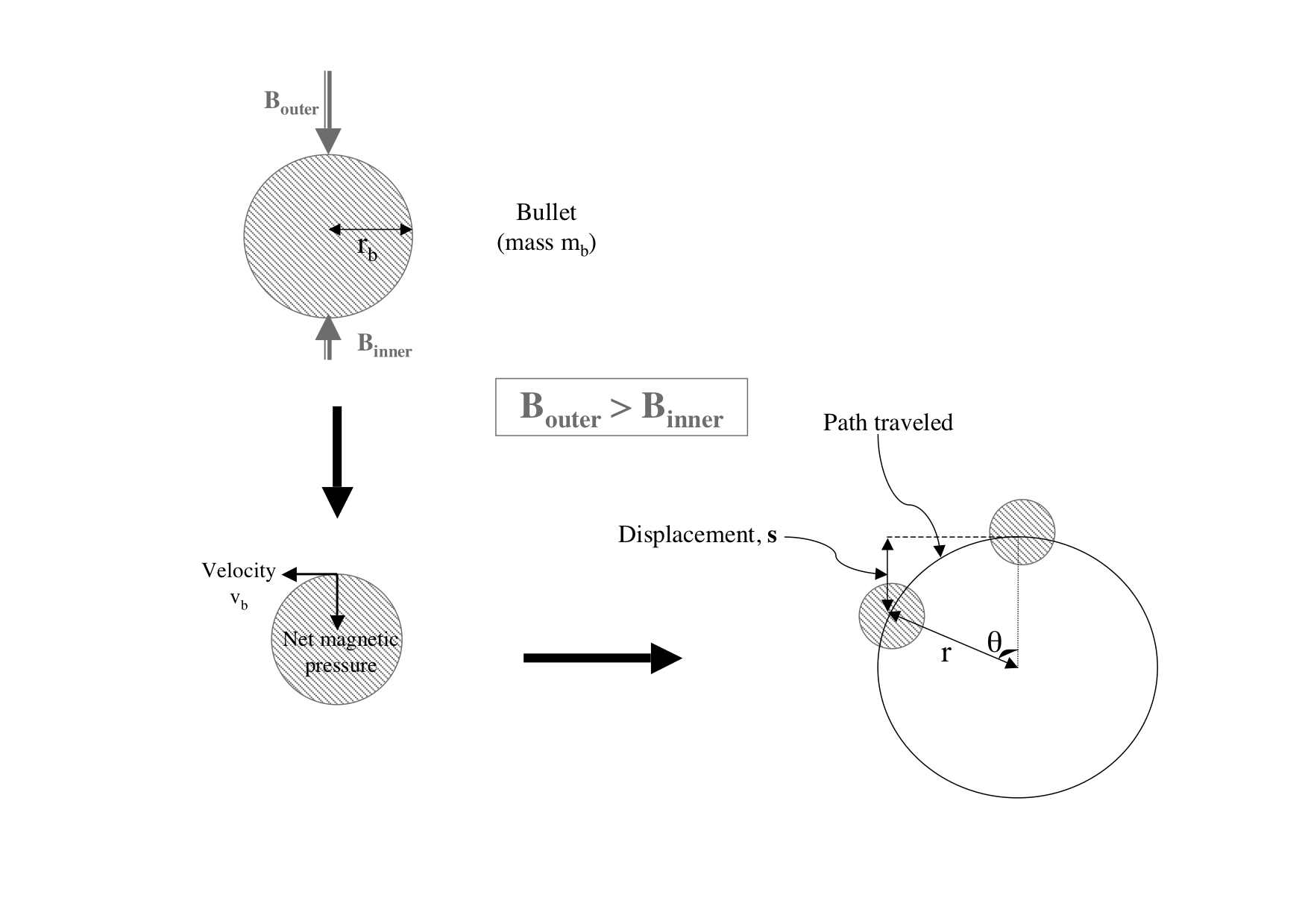}
 \caption{Cartoon illustrating the model adopted for the deflection in a magnetic field for an individual bullet of mass m$\rm_{b}$ and radius r$\rm_{b}$. 
The bullets which create the shocks are assumed to experience an excess magnetic pressure such that B$_{outer}$ $\ge$ B$_{inner}$ (top left). The bullets move with a velocity v$\rm_{b}$ along a curved path due to the excess magnetic pressure which they thereby feel (bottom left). The figure bottom right shows the displacement in the plane of the sky, \textit{s}, of a bullet between that arising from a rectilinear motion in the absence of a net magnetic pressure and the curved path in the presence of a magnetic pressure. The angle $\rm\theta$, see the derivation of Eq.~\ref{eq2}, is also illustrated. 
 }
             \label{Fig.cartoon}%
    \end{figure*}

For simplicity, imagine a magnetic field whose structure is a continuous small segment of the surface of a sphere. The true point of ejection, which we take to be the BN, I (n) centre of encounter, lies at some position within the sphere defining the segment. The segment is taken to be convex in our direction with the field increasing with radius. As a mass of ejected material encounters this field the differential magnetic pressure across any one mass of material will cause its path to follow a continuous geodesic, moving either northwest (Peak 1) or southeast (Peak 2) showing curvature both in the plane-of-the-sky and in a direction away from the observer, that is, the line-of-sight. Both motions will have the same radius of curvature. If we take tangents to these paths, equivalent to taking rectilinear trajectories, these tangents will meet at an apparent origin north of the true origin of the explosion projected onto the plane-of-the-sky. The present positions will also be bent away from the observer. 

We now seek to show that the displacement of the apparent origin from the BN, I (n) centre of encounter can be 6$\arcsec$ - 7$\arcsec$ given the known or likely properties of the system, and further that the angle at which the outflows are bent away from the observer is $\sim$15$\degr$ under the same set of physical conditions. We derive first some simple equations which govern the motion of the ejecta. If the difference between the magnetic field on the outer side of one of the ejecta and on the inner is $\Delta$B, with B$\rm_{outer}$ $>$ B$\rm_{inner}$ and the material effectively presents a surface of radius r$\rm_{b}$, then the net force sustaining the circular geodesic path of radius r will be $\pi$r$\rm_{b}\rm^{2}$$\Delta$B$\rm^{2}$/2$\rm \mu\rm_{0}$ where $\pi$r$\rm_{b}\rm^{2}$$\Delta$B$\rm^{2}$/2$\rm \mu\rm_{0}$~=~m$\rm_{b}$v$\rm_{b}\rm^{2}$/r where m$\rm_{b}$ is the mass of any bullet. Thus,

\begin{equation}
\label{eq1}
    \rm
       r =
\frac{2\mu_{0}m_{b}v^{2}}{\Delta B^{2}\pi r_{b}^{2}}. \\
    \end{equation}

The form of Eq. \ref{eq1} is essentially the same as that derived in \citet{Demichev:1965} given that the gradient of the field is linear within the mass of material. 

If the path sweeps out an arc covering an angle $\rm\theta$ then the mass will deviate from an extrapolated position based on a straight line trajectory to its true position by a distance, s, related to $\rm\theta$ through $\cos$($\rm\theta$) = (1-s/r). Thus for small s/r we find $\rm\theta$~$\sim$~(2s/r)$^{1/2}$. The distance travelled is $\rm\theta$r and therefore the time travelled is $\rm\theta$r/v given ballistic motion. Hence the time of travel, t, is given by 

\begin{equation}
\label{eq2}
    \rm
       t =
\frac{2(\mu_{0}m_{b}s/\pi)^{1/2}}{\Delta Br_{b}}. \\
    \end{equation}
%

A cartoon illustrating this model is shown in Fig. \ref{Fig.cartoon}. 

If we make the simplifying assumption that projectiles tend to be clustered around values of similar dimension and density it is evident that the displacement, s, will be similar for all ejecta. It is clear from Fig. \ref{Fig.Ringt0} that there are some outliers and this can be attributed to a breakdown of this assumption, if not to other over-simplified aspects of the model such as a  single value of net magnetic field.

The time t is taken to be 720~years from the present work. This may be compared with the figures of 500 and 560~years reported in \citet{Bally:2011} and \citet{Goddi:2011} respectively. We take the typical radius of a bullet to be 1.0$\arcsec$, estimated from the size of the perturbation created through the passage of the shock and the mass $m\rm_{b}$ to be a 5$\times$10$\rm^{-4}$~M${\sun}$ \citep{Burton:1997,Chrysostomou:1997}. Note that values of masses and velocities are consistent with the total energy of the outflow of $\sim$10$^{40}$~J \citep[e.g][]{Bally:2011,Chrysostomou:1997}. These parameters imply a density of n$\rm_{H}$ $+$ 2n$\rm_{H_{2}}$ $\sim$ 6$\times$10$\rm^{6}$~cm$^{-3}$, taking account of He, comparable with estimates that may be found for example in \citet{Tang:2010} and consistent according to those authors with a magnetic field of $\geq$9 mG. The observed displacement between the apparent centre and the actual centre of explosion is $\sim$4$\times$10$\rm^{14}$~m (6.5$\arcsec$). From Eq. \ref{eq2}, it follows that $\Delta$B would have to assume the value of $\sim$6~mG to achieve this displacement.

As mentioned above ejecta will also be bent away from the observer by the magnetic pressure engendered by $\Delta$B. Simple geometrical considerations, given motion on a sphere, show that the angle at which ejecta are deflected is given by $\tan\rm^{-1}$(s/d) where d is the distance that ejecta would have traveled from the apparent explosion centre in the absence of any magnetic deflection. Using values from above, it follows that this angle equals 14$\degr$. This implies an angle of 152$\degr$ between the flows, agreeing very well with the albeit approximate value of 150$\degr$ shown in Fig. \ref{Fig.angleP1P2}. Thus the magnetic deflection model has the advantage of offering an explanation for both the problem of the apparent lack of coincidence of the origin of the outflowing bullets with the BN, I (n) encounter centre and also the deflection backwards out of the plane of the sky of Peaks 1 and 2. 

Data in \citet{Houde:2004,Vaillancourt:2008,Tang:2010} demonstrate the presence of a highly ordered magnetic field covering the region of interest in Fig. \ref{Figure1} at a position angle $\sim$130$\degr$  east of north. With regard to absolute values of the field, \citet{Tang:2010} argue that a (tangential) magnetic field of anything between 3 and 30~mG may be required to confine ejecta. These considerations suggest that an ordered field of several mG is appropriate for the gas through which the ejecta travel. However within the cavity created by expulsion of material due to the explosion, the magnetic field would be likely to be very low since the field would be expelled with the ionized material. Note that a marked absence of material is observed in the region around the ejection centre. This is illustrated for example by the absence of dust emission around the explosion centre in continuum maps in~\citet{Tang:2010}. It is also illustrated by the absence of H$\rm_{2}$ emission and warm dust emission in images in \citet{Lacombe:2004} where H$_2$ emission should otherwise be generated through the high UV flux from the Trapezium in this region were there material present at densities of a few times 10$^{3}$~cm$^{-3}$ or higher. 
Thus we propose a fragment of a spherical cavity with material outside the cavity maintaining a field of several mG and a small field within the cavity. This would give rise to the magnetic gradient which is a necessary ingredient of the above model for deflection of the trajectories of ejecta. 
In this connection magnetic fields derived from OH maser Zeeman splittings vary very considerably -- between -3.5~mG and +16.3~mG -- within the central arcsecond around the position of radio source I and do not show a clear pattern \citep{Cohen:2006}. These fields are local to each maser spot and may be wound up by local turbulent eddies~\citep{Field:1982}.

Notwithstanding the broad success of this magnetic deflection model we have omitted to include magnetic deflection in tracing out the motions of the bullets in any of the relevant figures in this paper. This choice has been made in order to maintain the results as strongly observational based as feasible. The introduction of  magnetic deflection would necessitate assumptions about the mass, magnetic field and bullet density distributions.



\section{Conclusions}
\label{sec:conclusions}

Combining proper motion and radial velocity data it has proved possible to create 3D images of the structure of \object{OMC-1} in vibrationally excited H$\rm_{2}$ in the inner 0.1~pc. These graphic images are based directly on observational data and require few assumptions for their creation. The most significant assumptions are that the H$\rm_{2}$ bullets lie in the plane of the sky at the epoch of the explosion and that the motion is ballistic. This latter assumption may readily be relaxed without any qualitative implications for our understanding.

The magnetic deflection model introduced here, simple as it is, removes any lingering doubt that the outflow was in some way triggered by the conjunction of BN, I (n) at an epoch between 500 and 700 years in the past. This model is able to reconcile the systematic departure of the apparent explosion centre, based on linear extrapolation of proper motions, with the position of the BN, I (n) encounter, using parameters which appear quite appropriate to the region of interest. 

Results presented here suggest that the explosion impacted on a pair of distorted arcs of molecular material forming a ring $\sim$15$\arcsec$ (6000~AU) in diameter. We also see how Peaks 1 and 2 are swept away from the observer into the plane of the sky by $\sim$15$\degr$. Images further reveal the disposition of the slow blue-shifted outflow, Region B, which is found to be angled to the south and west in a manner which does not appear to relate to the fast outflow but may be connected with the motion of I if it is the source of the slow outflow.

In conclusion, while discrepancies remain with regard to the details of the close encounter of massive bodies within the core of \object{OMC-1} and the overriding contributions to the mechanism which led to the explosion, a remarkably consistent phenomenological picture now emerges at least with regard to the fast outflow forming Peaks 1 and 2. 



\begin{acknowledgements}
We would like to acknowledge the support
of the Aarhus Centre for Atomic Physics (ACAP), funded by the Danish
Basic Research Foundation and also financial support from the Instrument Centre for Danish Astrophysics (IDA),
funded by the Danish National Science Committee (SNF, now FNU). J.L.L.
would like to acknowledge the support of the PCMI National Program, funded
by the French Centre National de la Recherche Scientifique (CNRS). We also
wish to thank the Directors and Staff of CFHT for making possible
observations reported in this paper and for the assistance rendered by E.
Le Coarer (Observatoire de Grenoble) in obtaining GriF data. We also wish to
thank G. Pineau Des For\^{e}ts and Lars E. Kristensen for valuable discussions about
the nature of shocks. Finally we thank the anonymous referee for many helpful comments.
\end{acknowledgements}


\bibliographystyle{aa}



\Online

\appendix 
\section{3D view of the \object{OMC-1} outflow}
\label{Appendix-movie}

The following figures allow the 3D reconstruction of the \object{OMC-1} outflow observed in the H$\rm_{2}$ S(1) v = 1--0 line from the expansion centre $\le$~720~years ago to the present time.
The material may be viewed from different vantage points in space. Complete animations may been obtained upon request to the authors (D.~F., H.~N. or C.~F.).

\onlfig{1}{
\begin{figure*}
\centering
\includegraphics[width=4cm,angle=0]{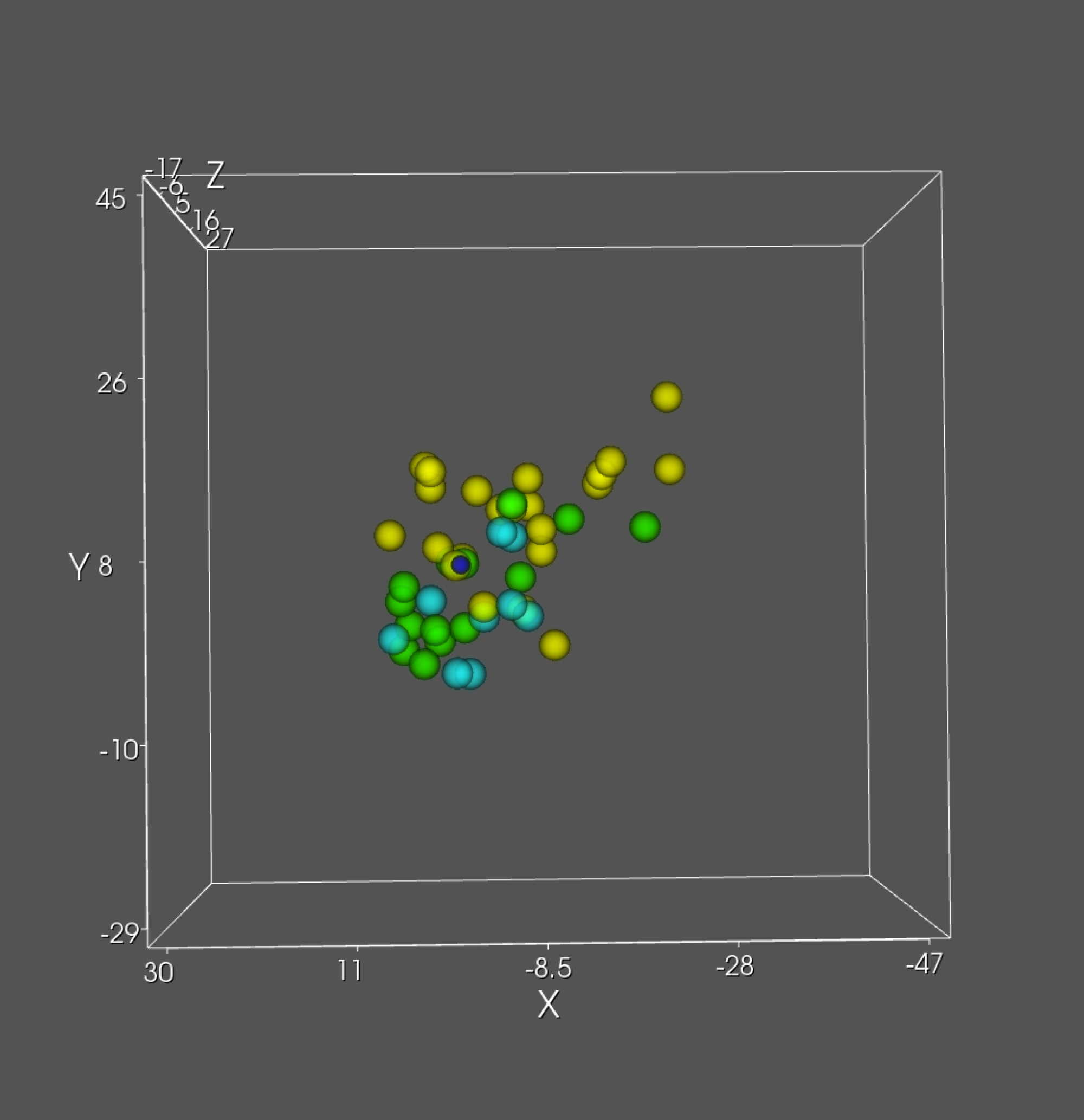}
\includegraphics[width=4cm,angle=0]{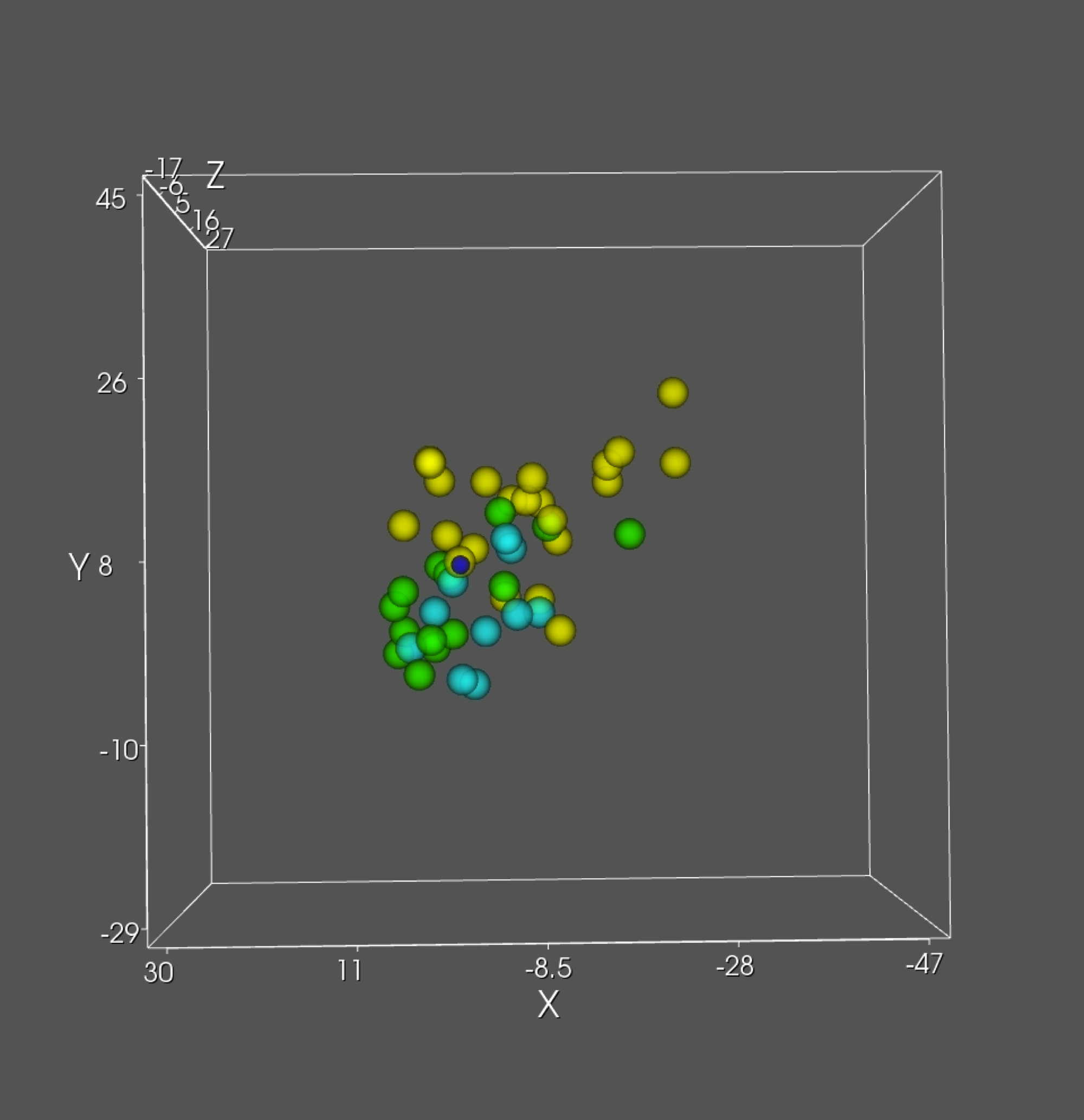}
\includegraphics[width=4cm,angle=0]{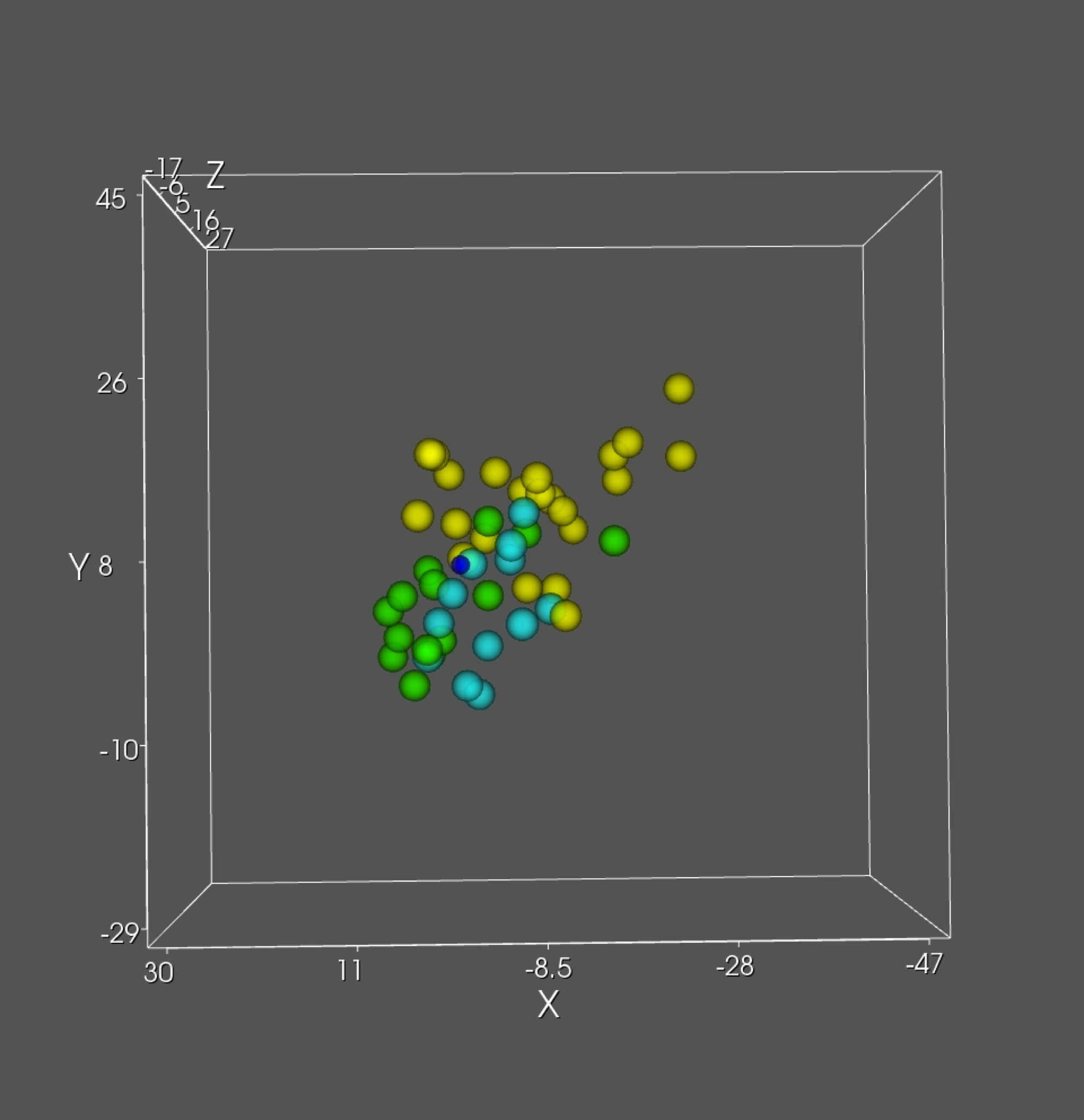}
\includegraphics[width=4cm,angle=0]{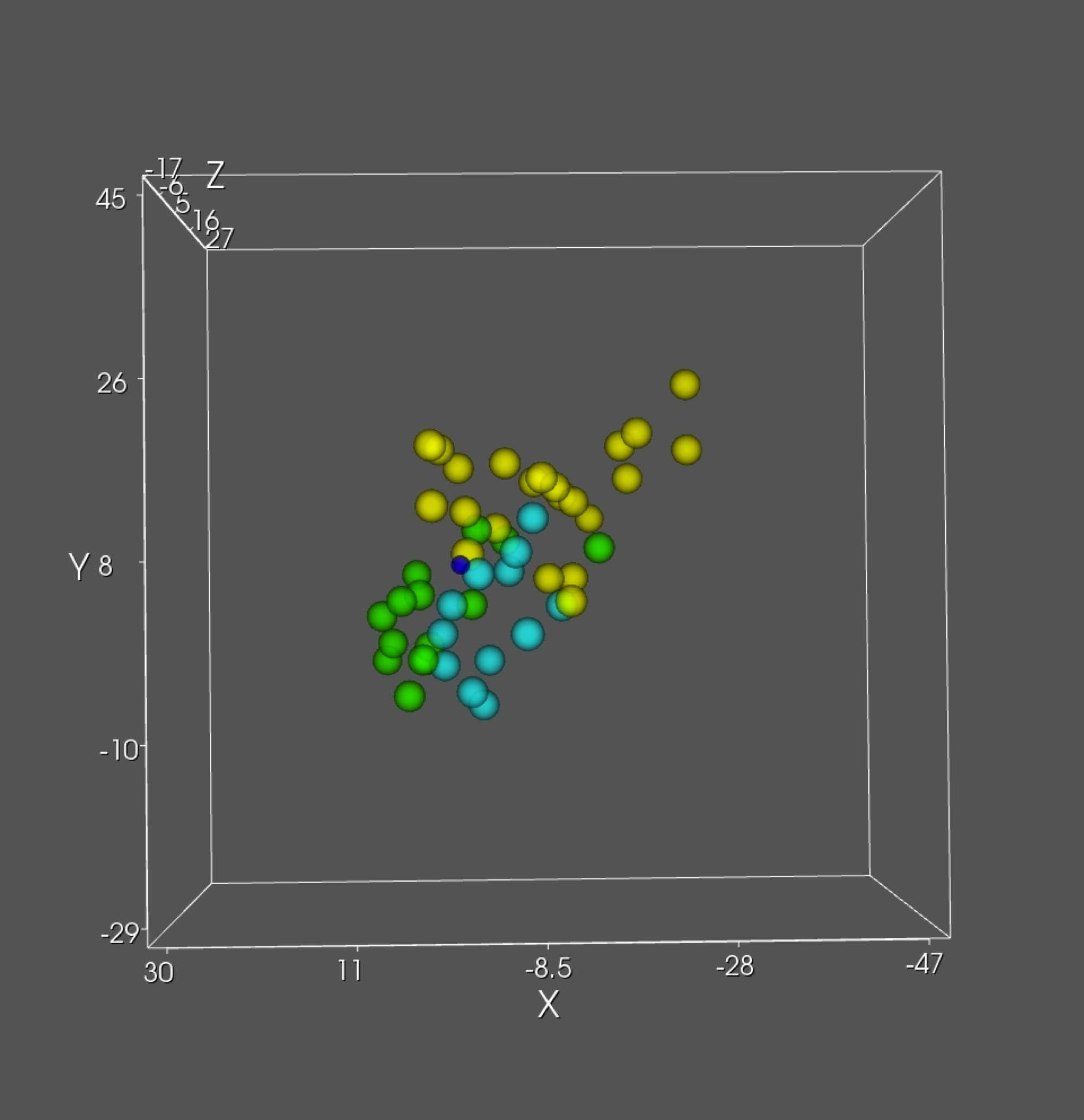}
\includegraphics[width=4cm,angle=0]{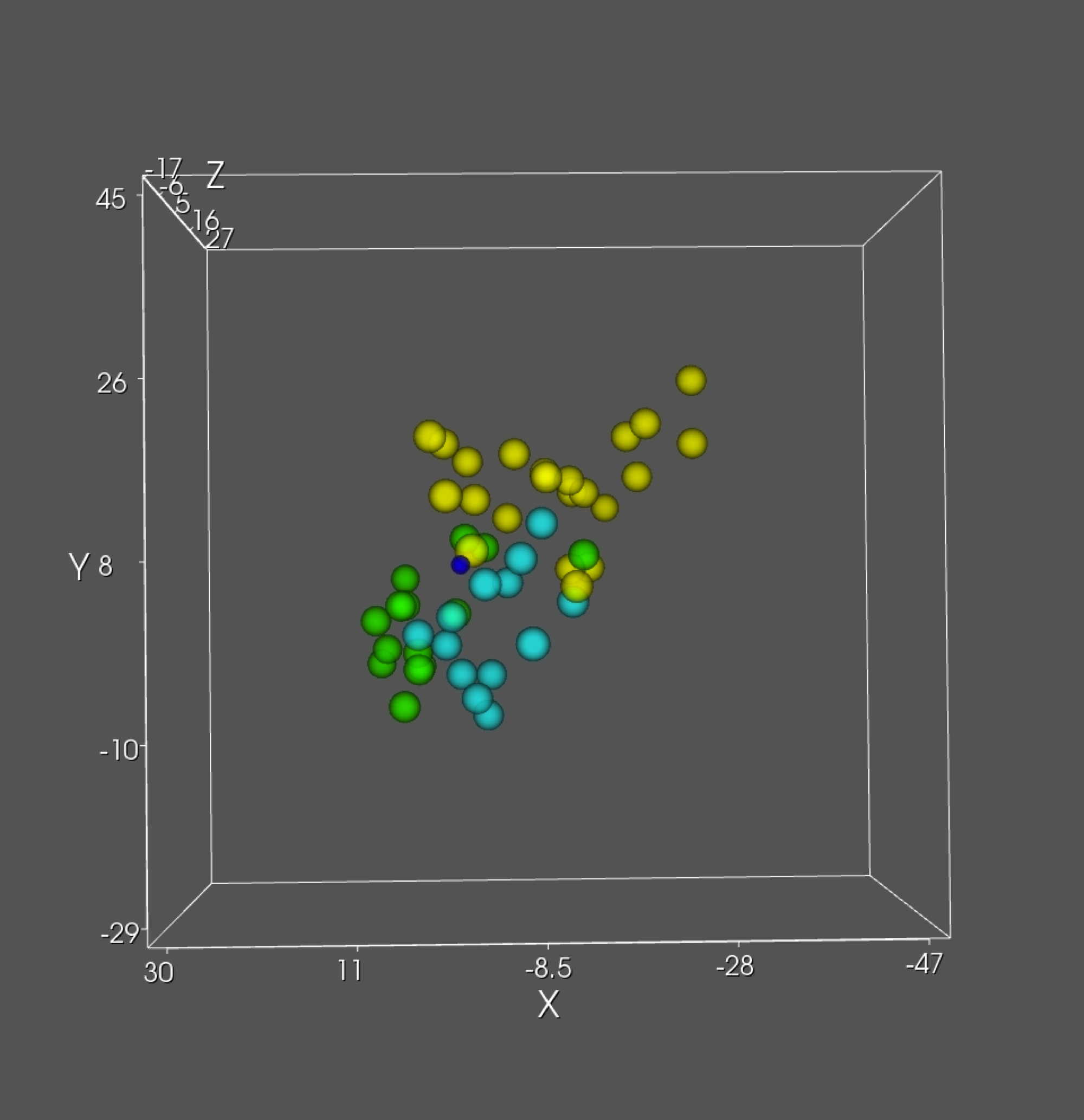}
\includegraphics[width=4cm,angle=0]{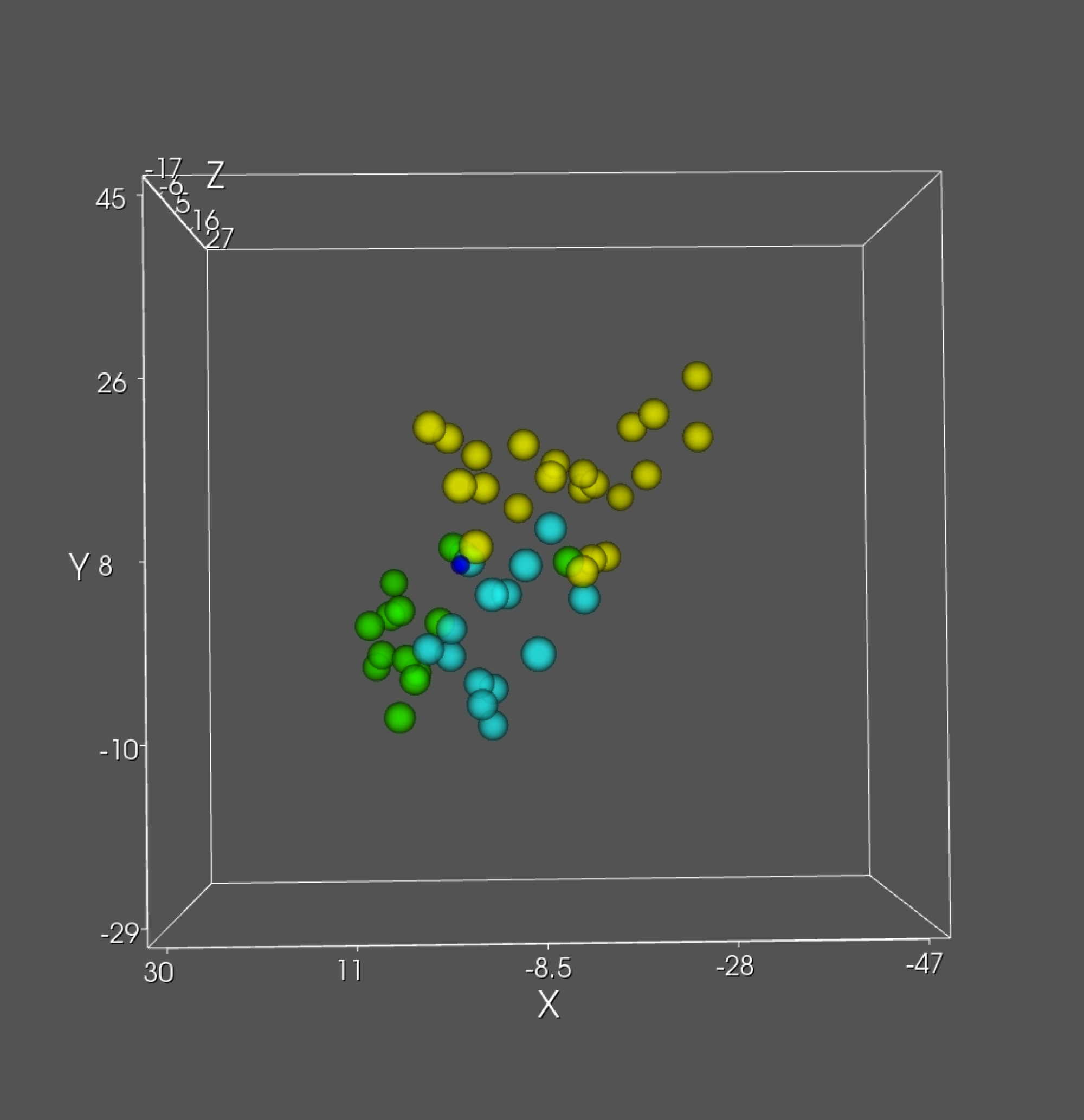}
\includegraphics[width=4cm,angle=0]{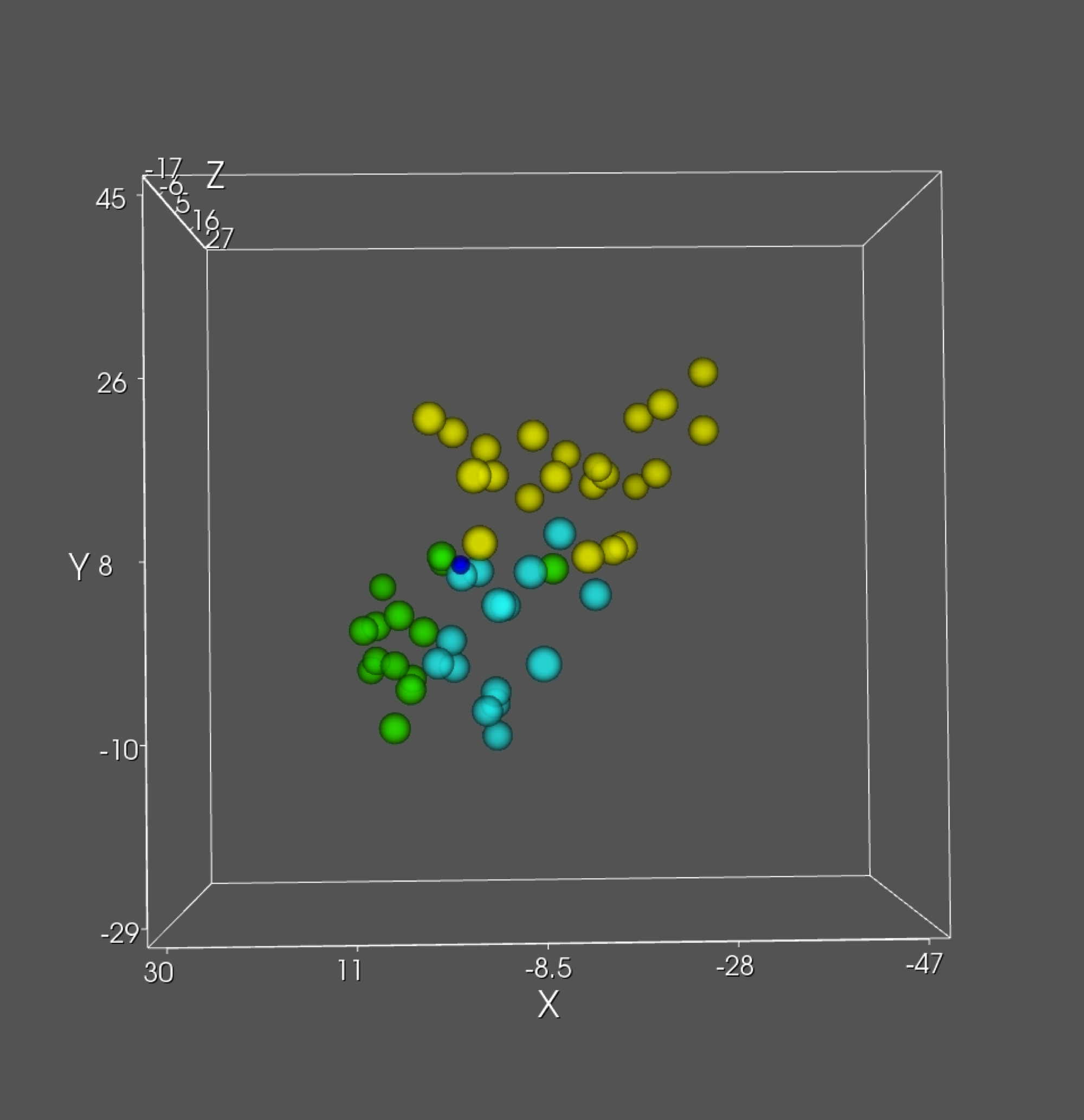}
\includegraphics[width=4cm,angle=0]{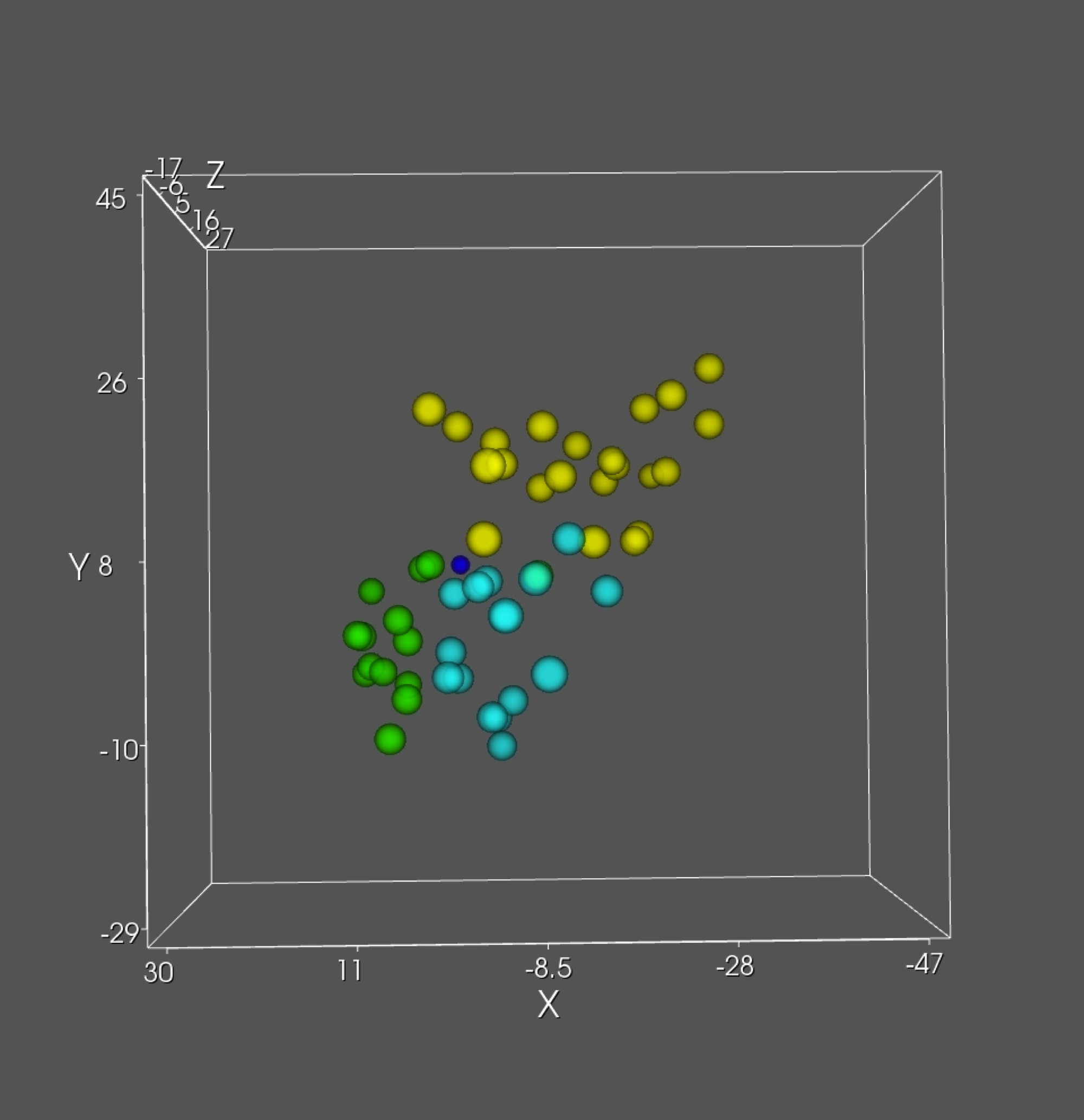}
\includegraphics[width=4cm,angle=0]{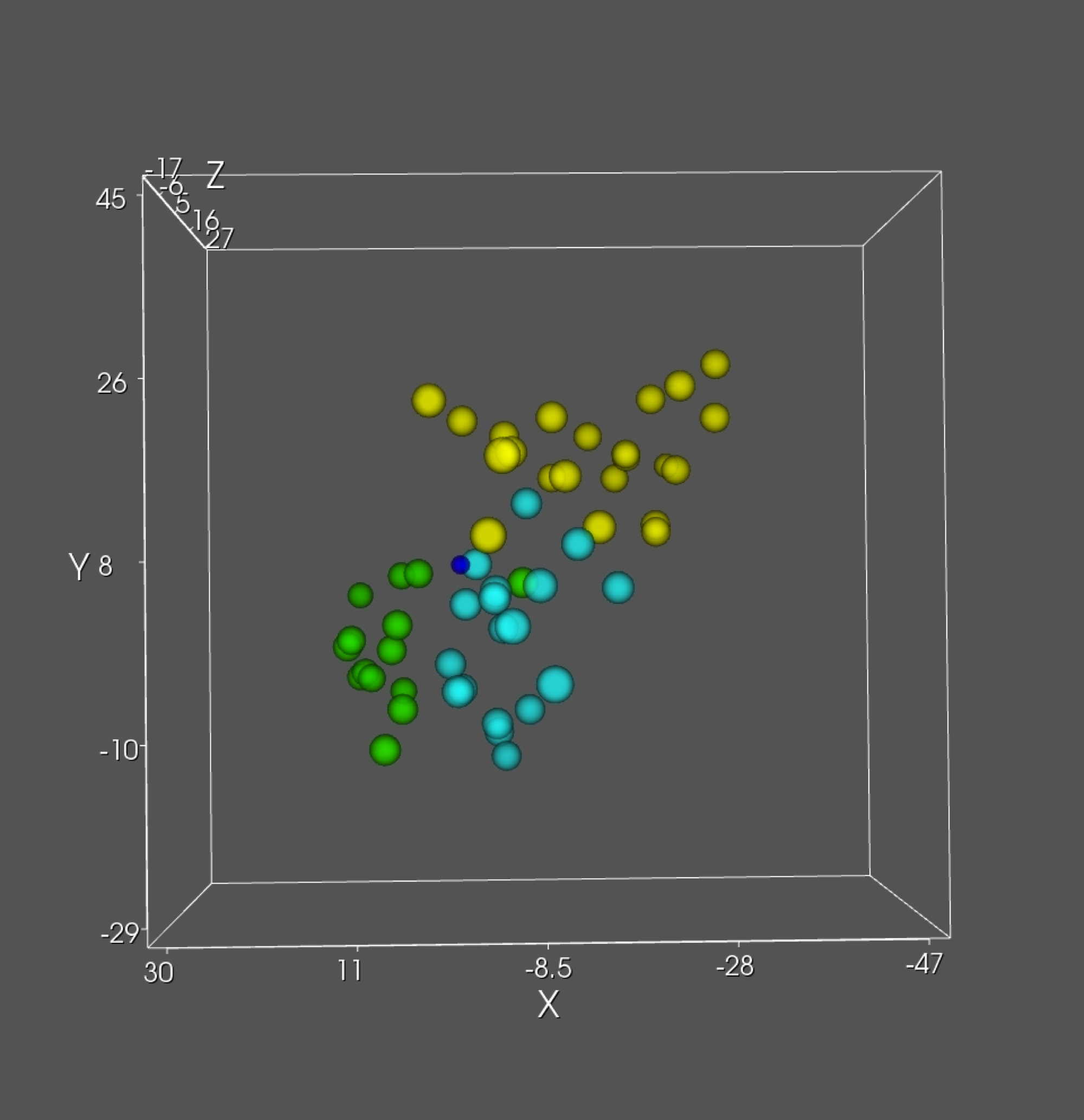}
\includegraphics[width=4cm,angle=0]{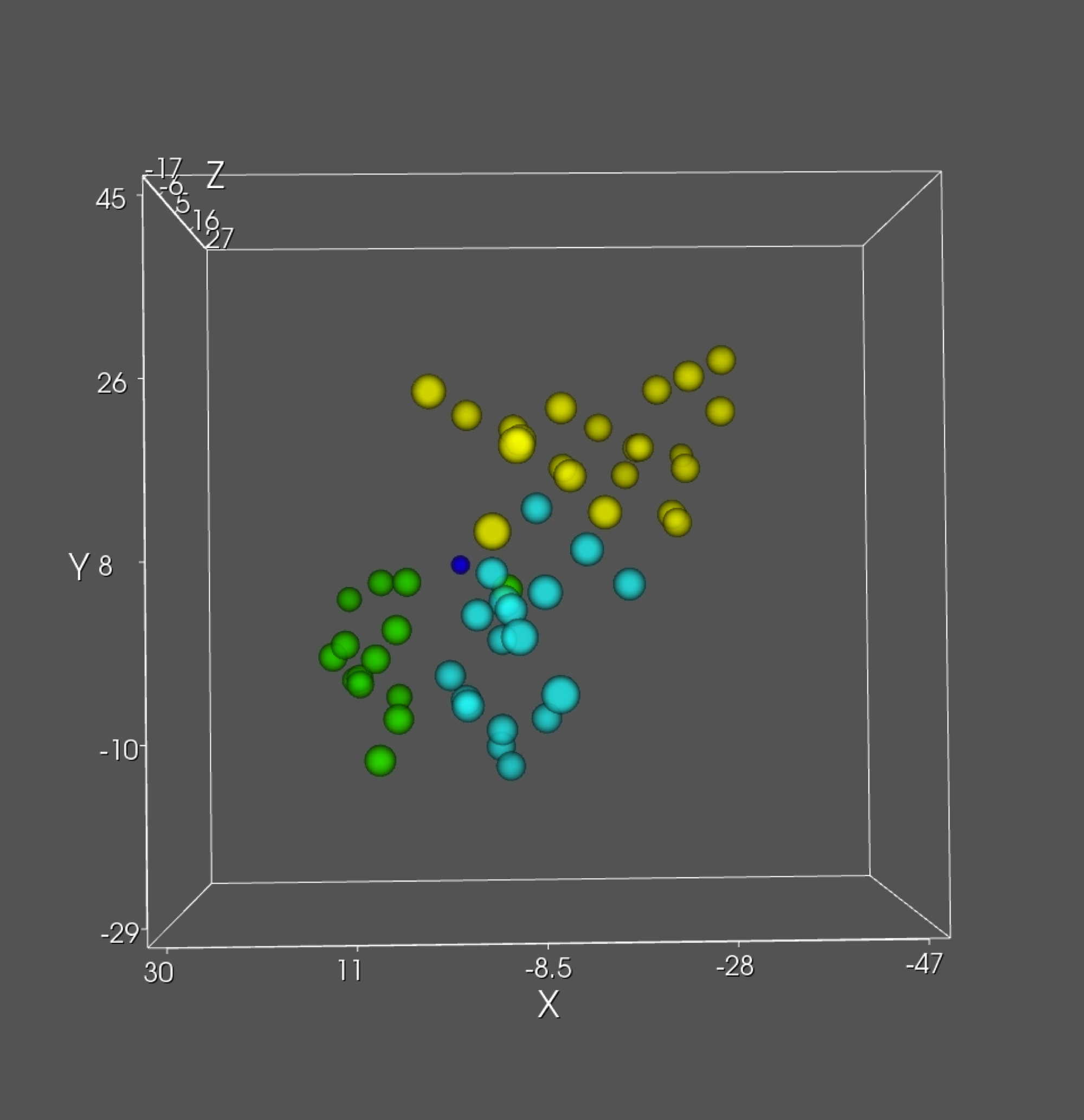}
\includegraphics[width=4cm,angle=0]{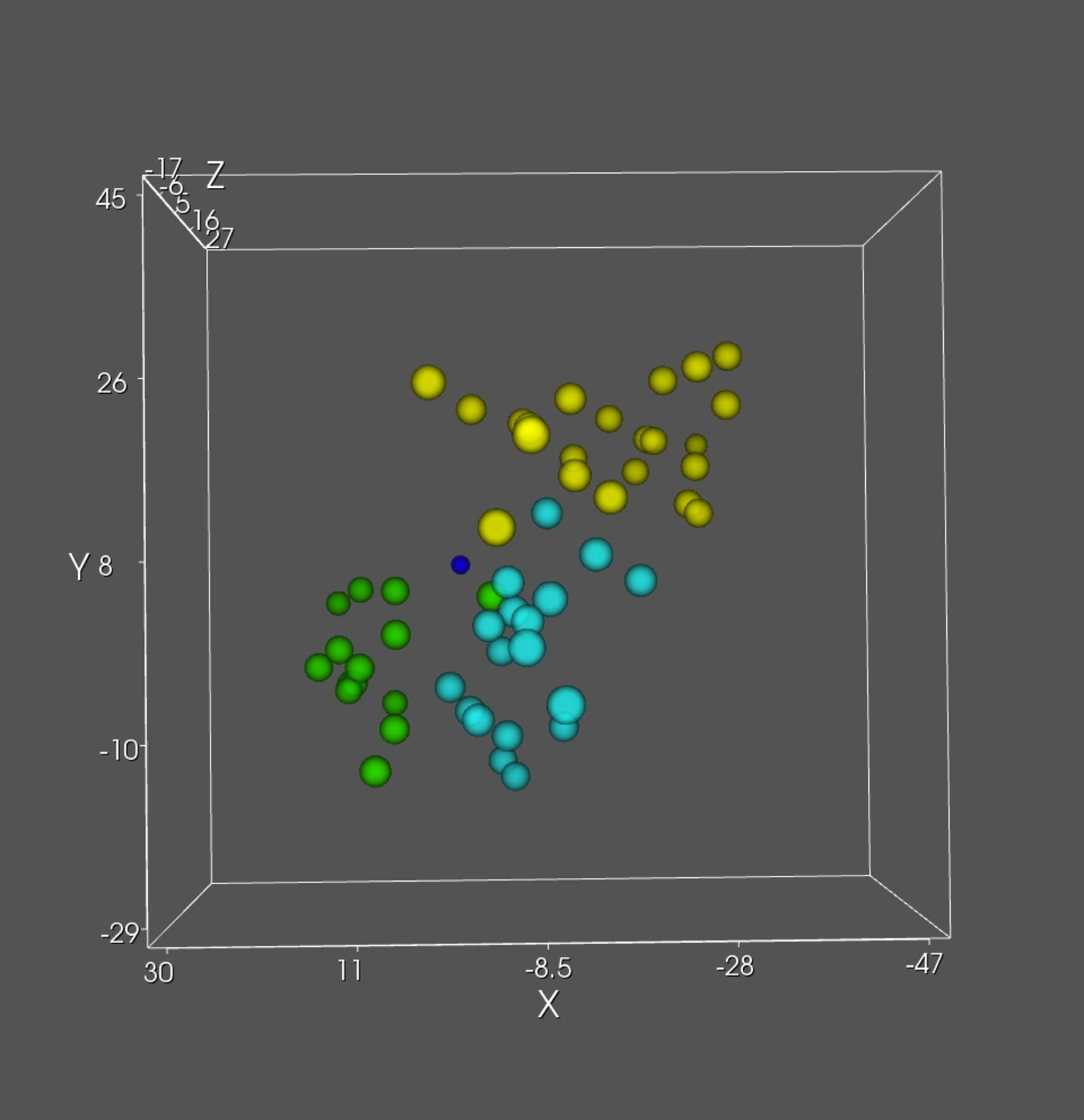}
\includegraphics[width=4cm,angle=0]{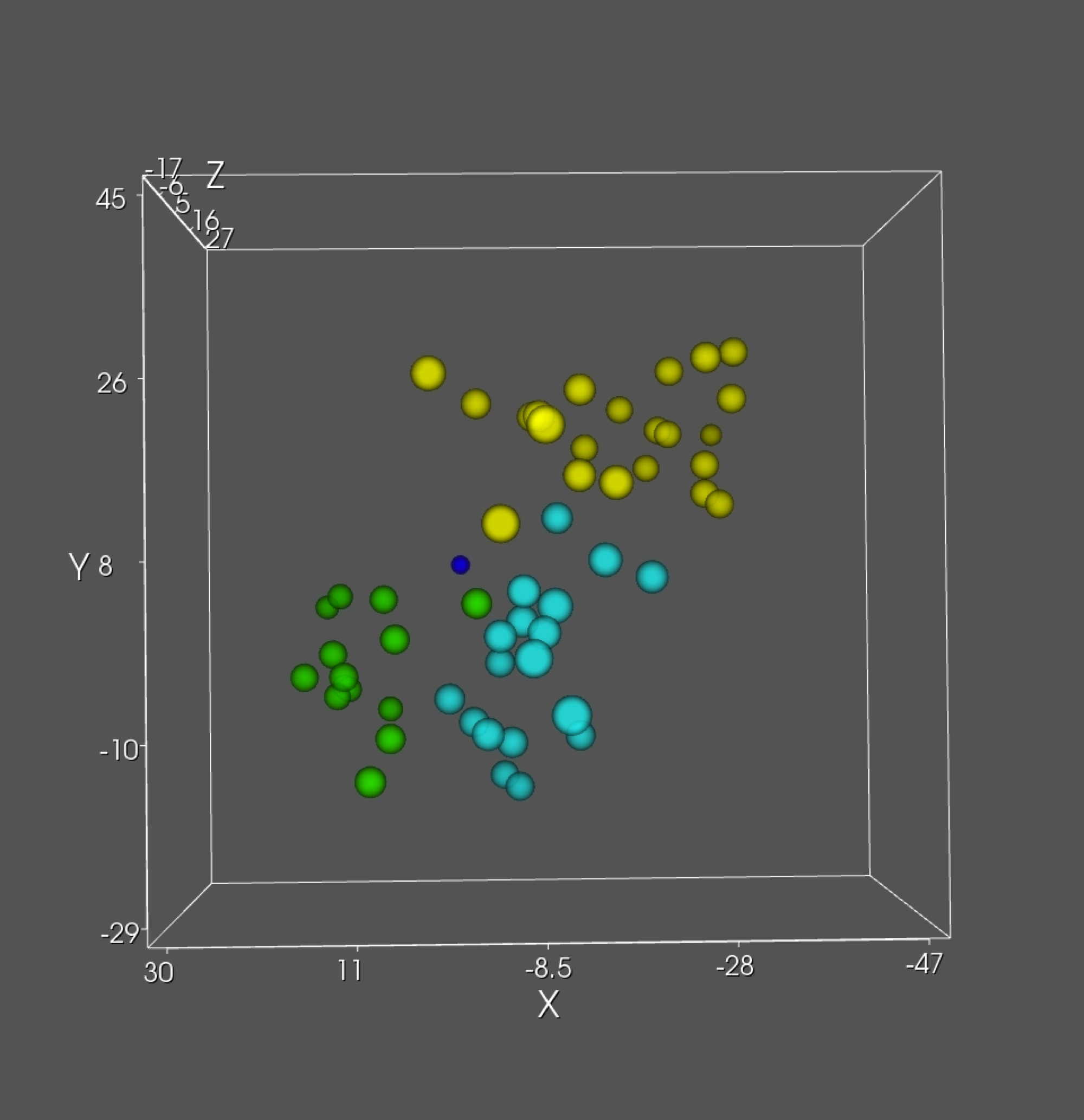}
\includegraphics[width=4cm,angle=0]{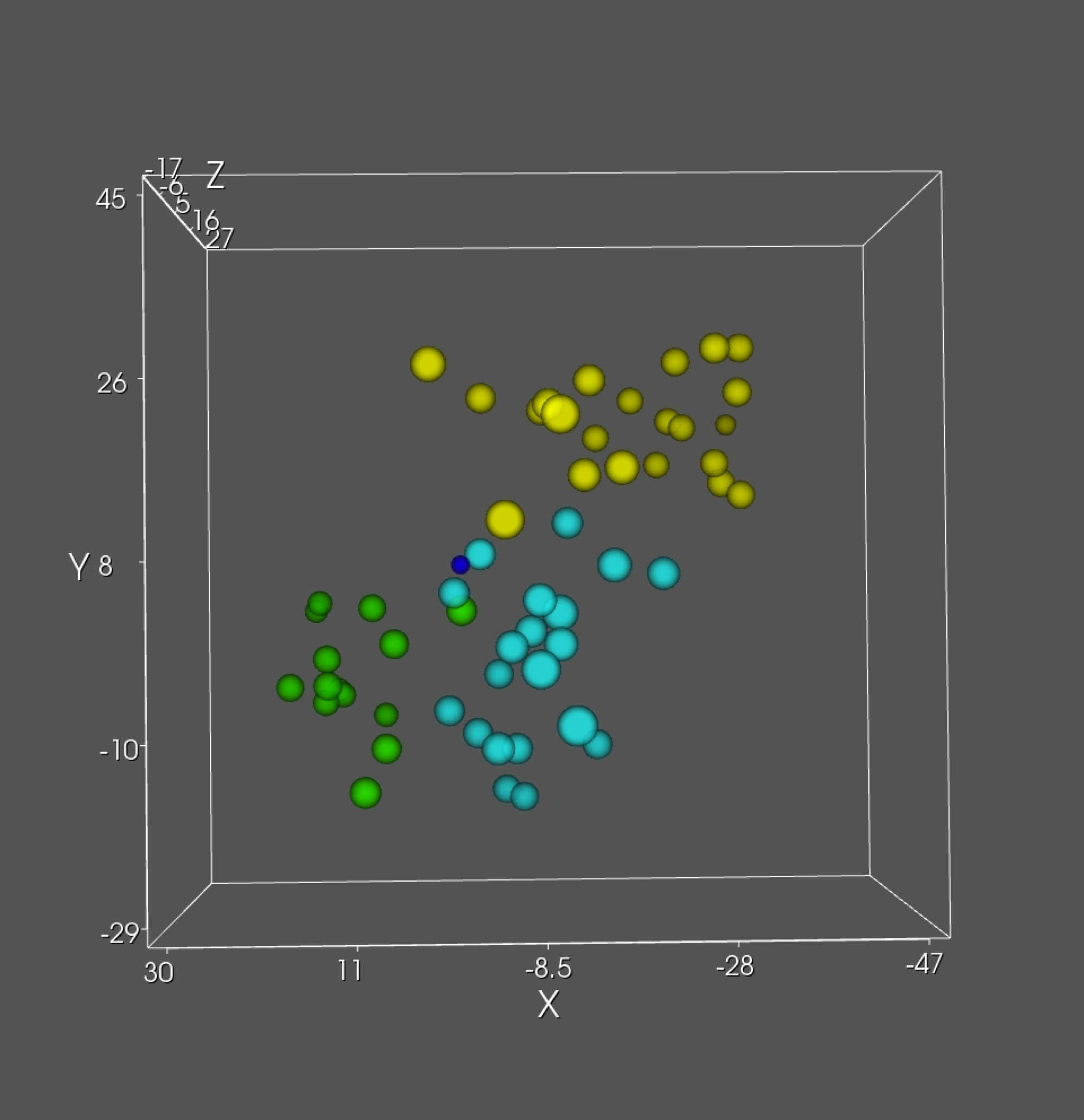}
\includegraphics[width=4cm,angle=0]{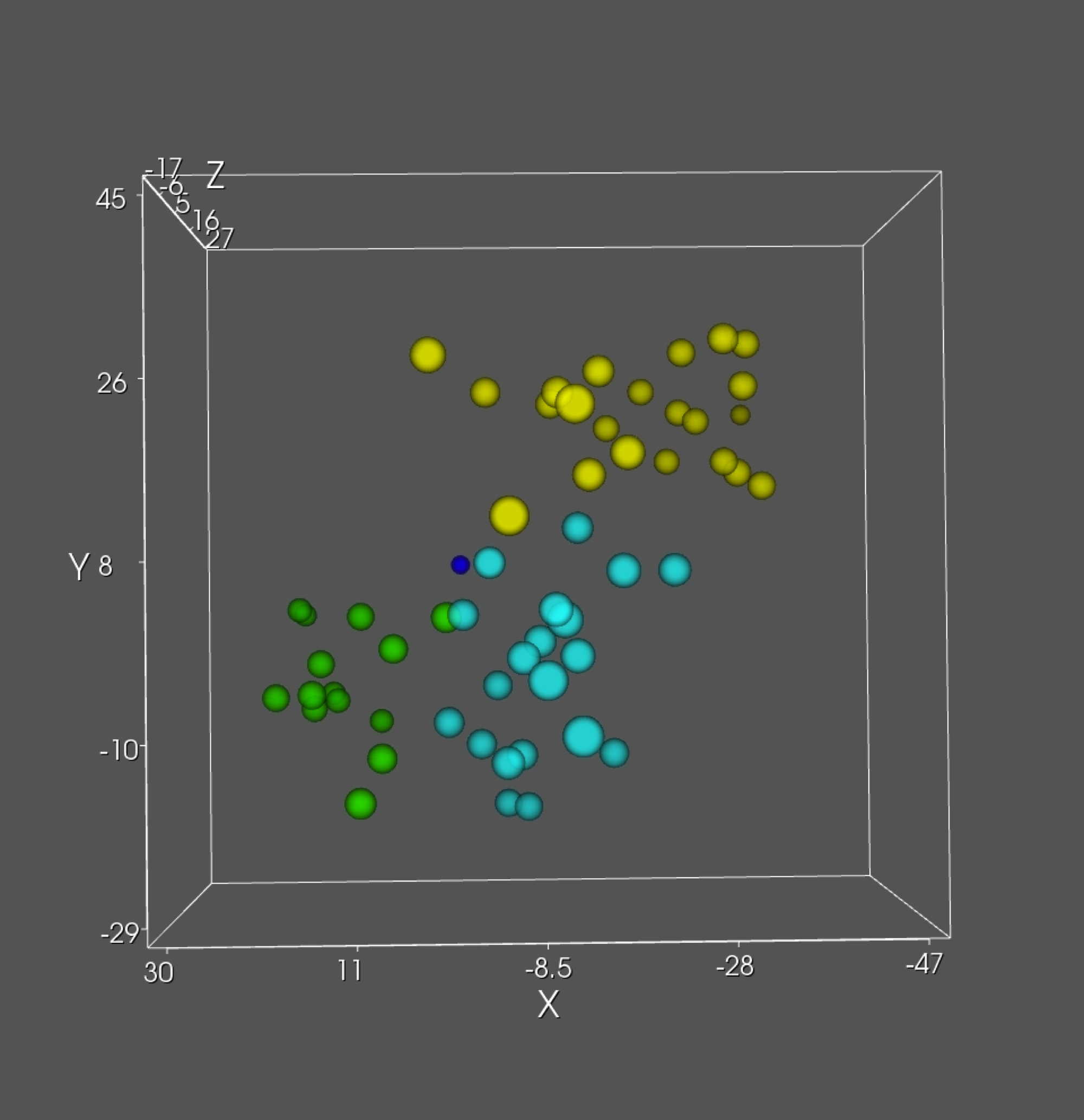}
\includegraphics[width=4cm,angle=0]{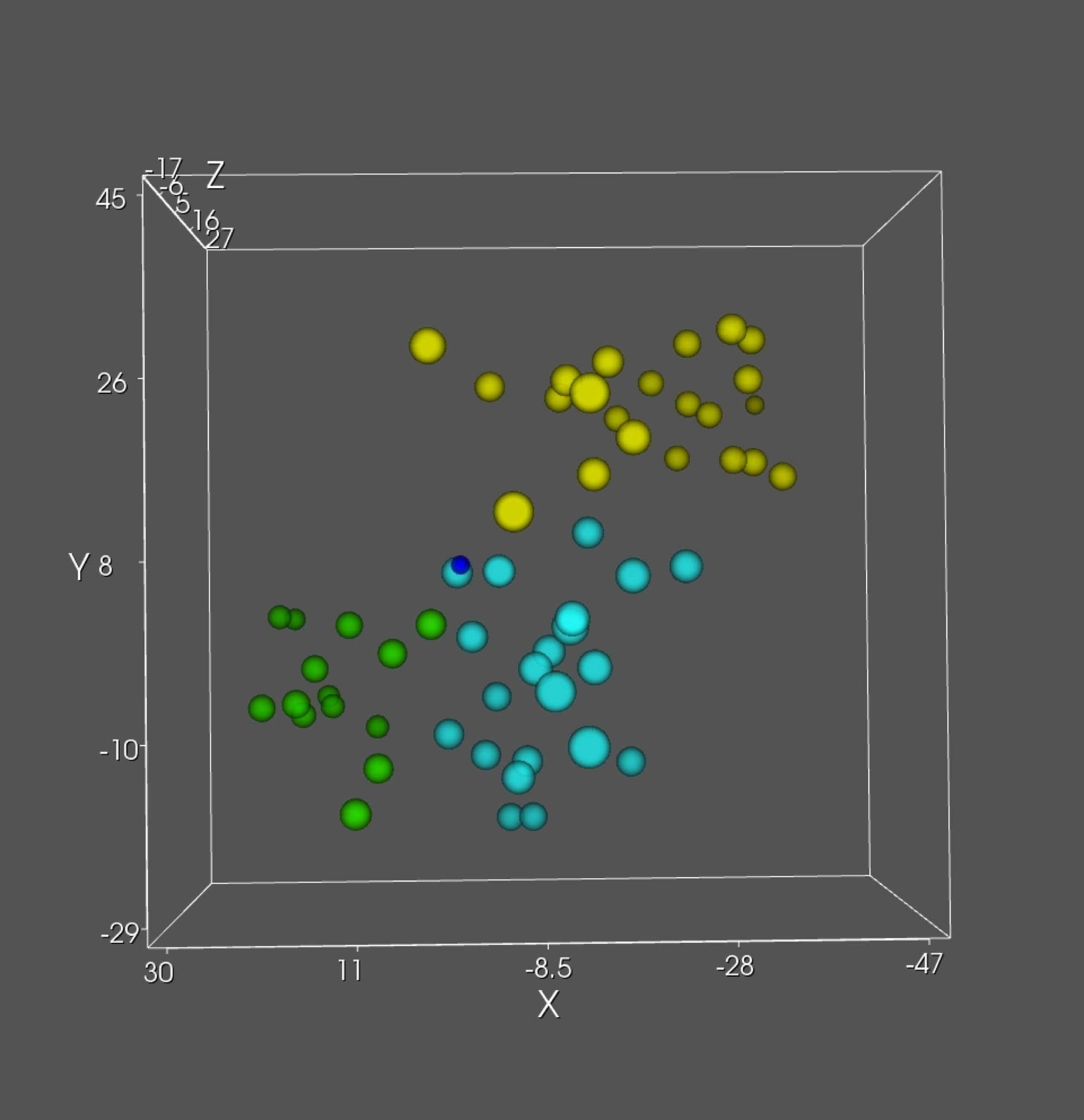}
\includegraphics[width=4cm,angle=0]{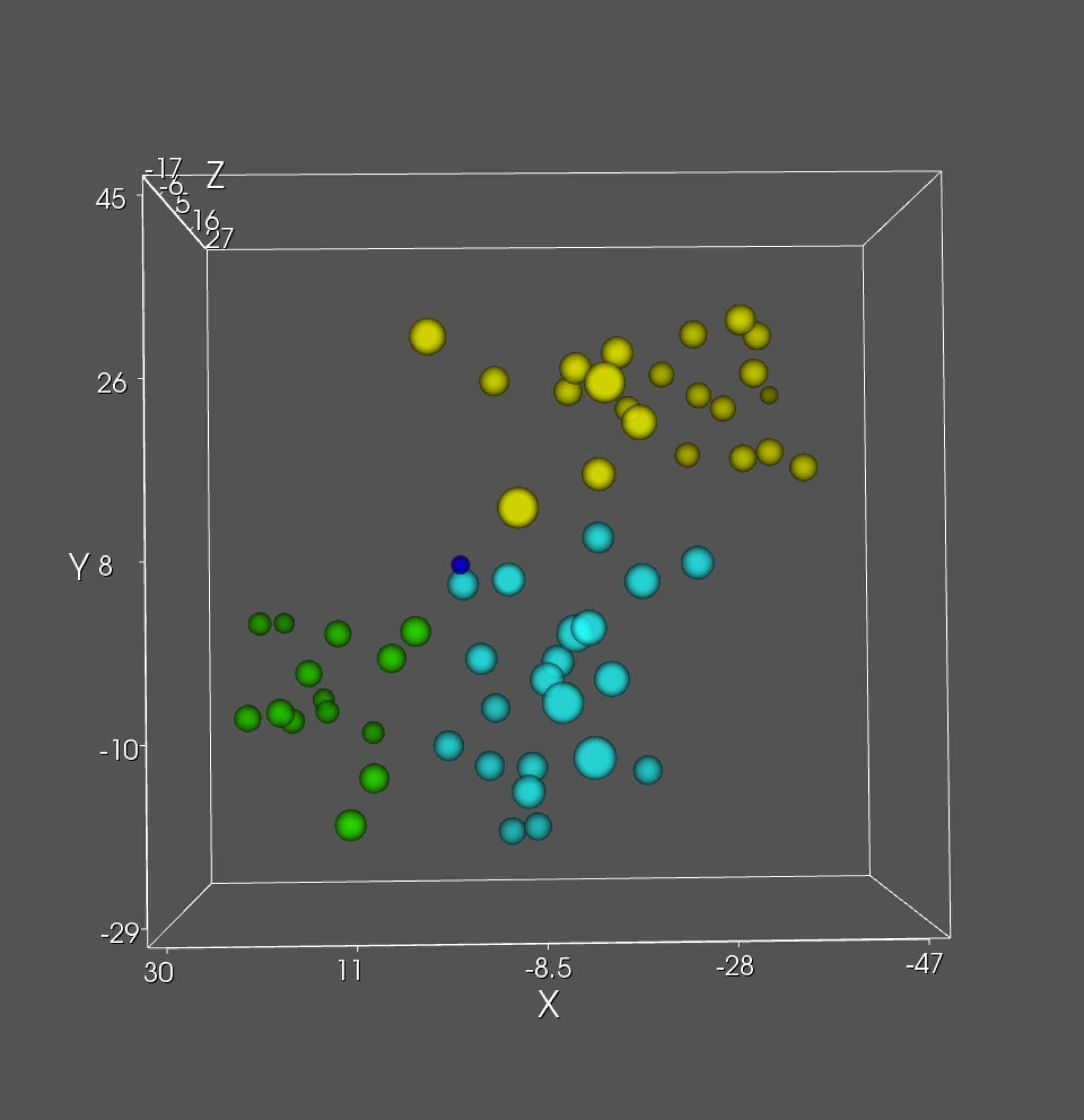}
\includegraphics[width=4cm,angle=0]{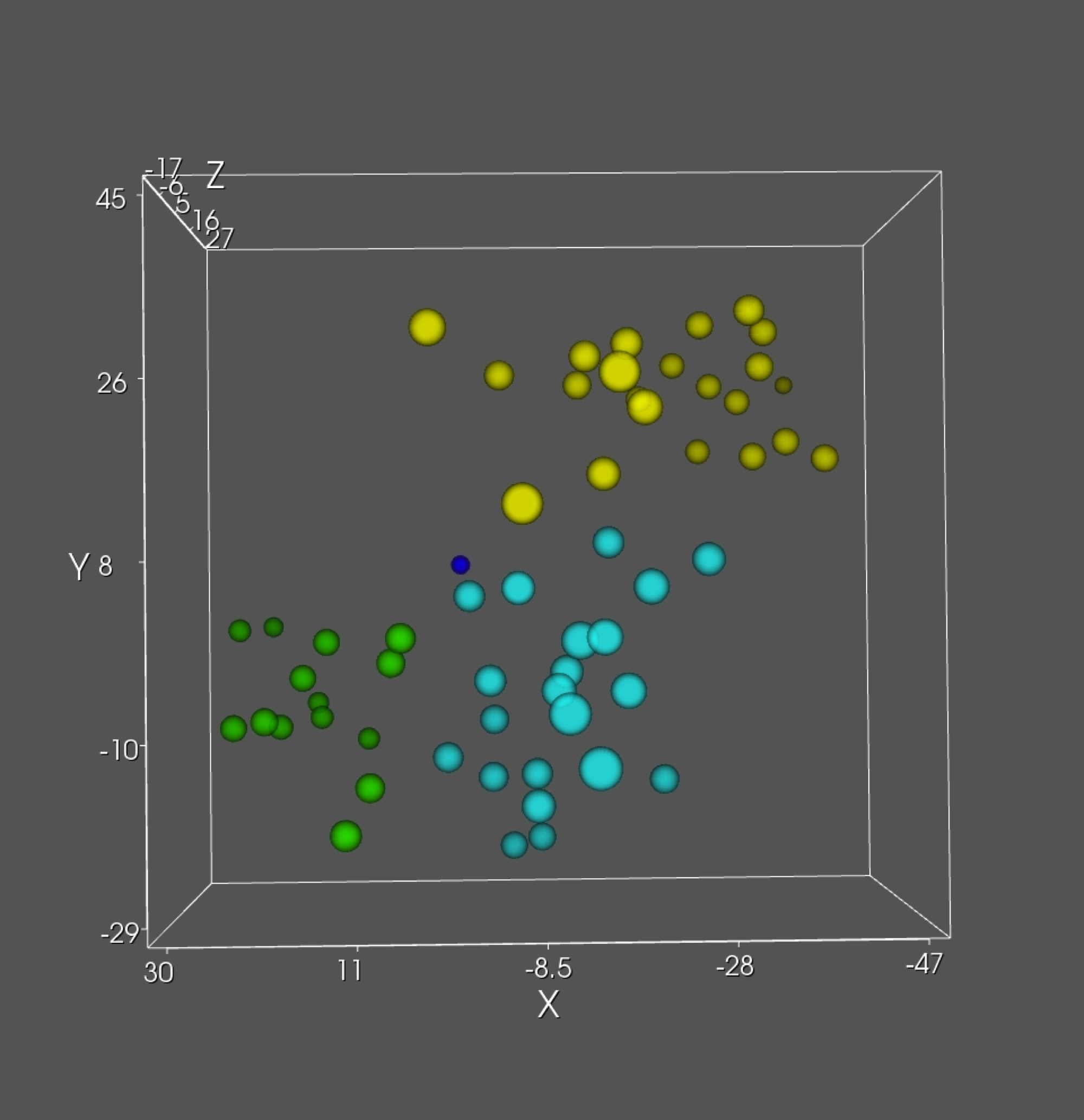}
\includegraphics[width=4cm,angle=0]{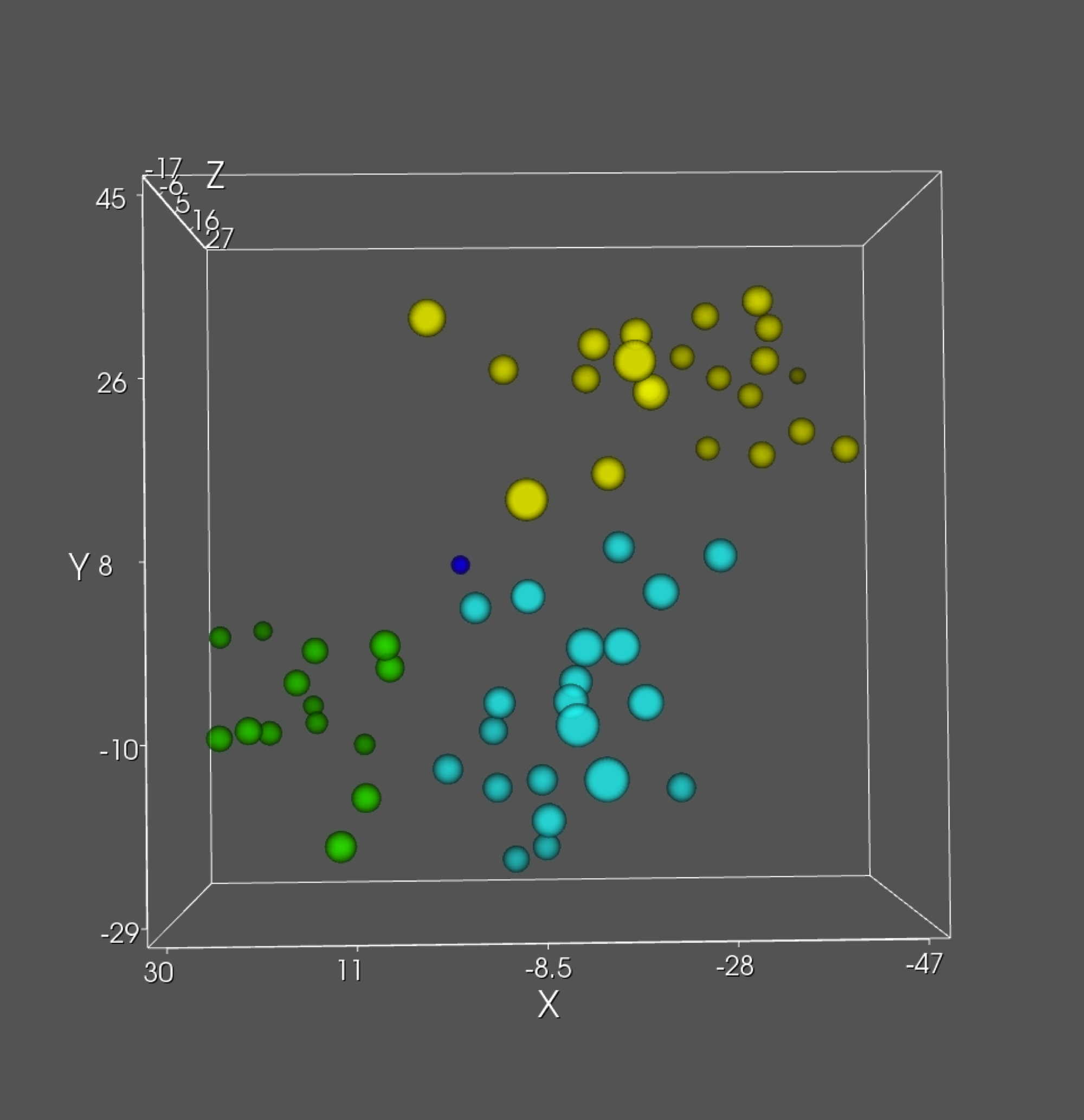}
\includegraphics[width=4cm,angle=0]{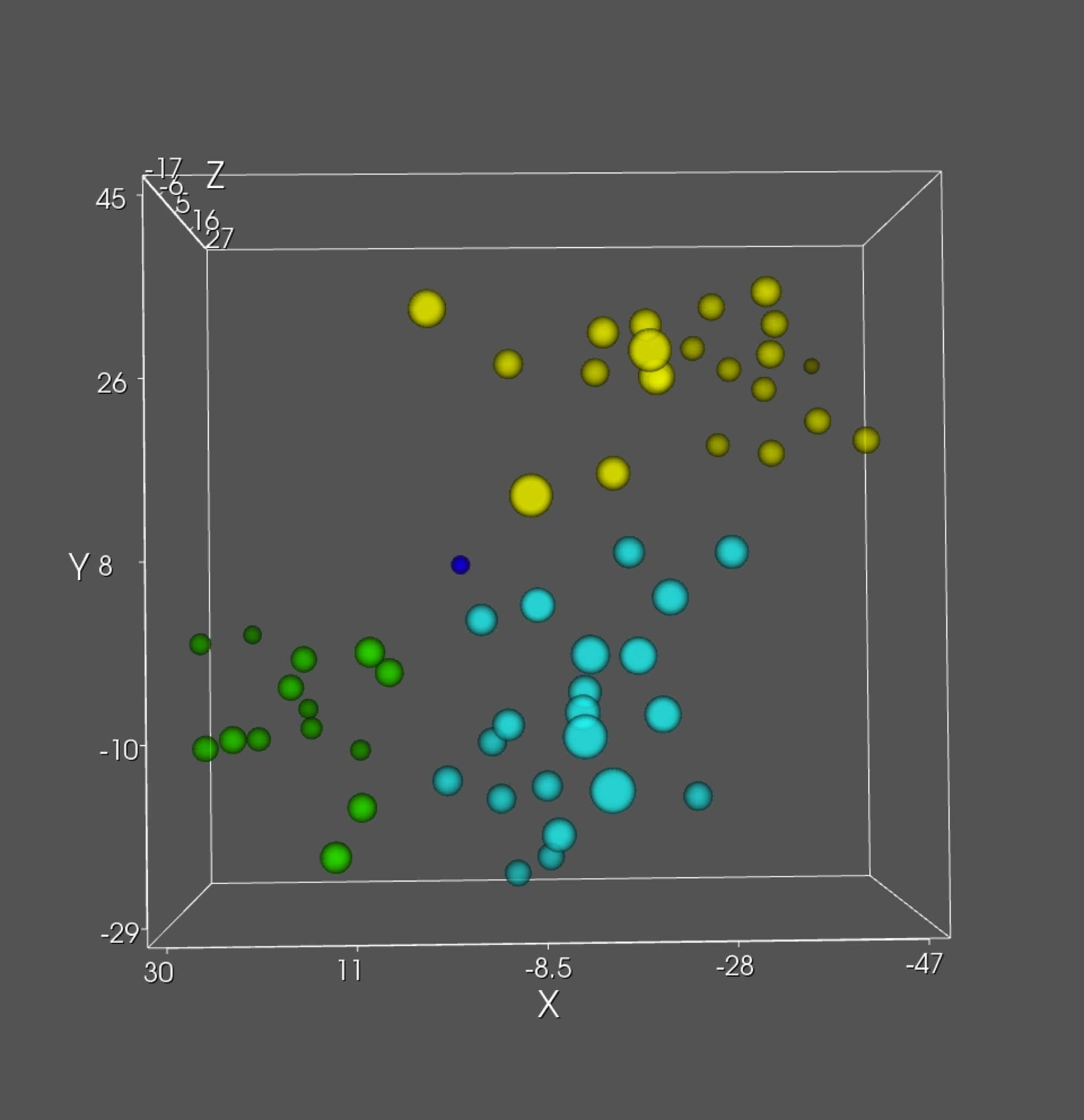}
    \caption{3D view of the OMC-1 outflow from the expansion centre $\le$~720~years ago to the present time. The blue dot is the computed expansion centre ($\alpha_{J2000}$ = 05$^{h}$35$^{m}$14$\fs$50, $\delta_{J2000}$ = -05$\degr$22$\arcmin$23$\farcs$00). Yellow, green and turquoise points show the disposition of the Peak 1, Peak 2 and Region B objects through time. The snapshots are at intervals of 40~years. xy is the plane of the sky. Notwithstanding the magnetic deflection model introduced in section 5, we have for simplicity and to avoid model dependence used rectilinear trajectories in this and all other relevant figures. }
      \label{3dmovreco} 
    \end{figure*}
    }
%

%
\end{document}